\documentstyle[aps]{revtex}

\newcommand{\fr}[2]{\frac{\displaystyle{#1}}{\displaystyle{#2}}}

\newcommand{\cA}{{\cal A}}

\newcommand{\mod}{\:{\rm mod}\:}

\newcommand{\p}{^{(\pm)}}
\newcommand{\m}{^{(\mp)}}
\newcommand{\uo}[1]{\underline{\omega}_{#1}}
\newcommand{\gpm}{g_n^{(\pm)}}
\newcommand{\gp}{{g_n^{(+)}}}
\newcommand{\gm}{{g_n^{(-)}}}
\newcommand{\Mprw}{{M_q^{\rm prw}}}

\input{psfig}

\begin{document}
\draft
\title{\bf Entropy Production in a Persistent Random Walk}
\author{T. Gilbert\thanks{Current address~: Laboratoire de Physique Th\'eorique
de la Mati\`ere Condens\'ee, Universit\'e Paris VII, B.~P.~7020, Place Jussieu,
75251 Paris Cedex.} and J. R. Dorfman}
\address{Department of Physics and Institute for Physical Science and
Technology\\University of Maryland, College Park, Maryland 20742}
\date{\today}
\maketitle

\begin{abstract}
We consider a one-dimensional persisent random walk viewed as a 
deterministic process with a form of time reversal symmetry. Particle 
reservoirs placed at both ends of the system induce a density current
which drives the system out of equilibrium. The phase space 
distribution is singular in the stationary state and has a cumulative
form expressed
in terms of generalized Takagi functions. The entropy production
rate is computed using the coarse-graining formalism of Gaspard, Gilbert and
Dorfman. In the continuum limit, we show that the value of the
entropy production rate is independent of the coarse-graining and
agrees with the phenomenological entropy production rate of 
irreversible thermodynamics.



\end{abstract}

\section{\label{I}Introduction}
%\eqlabel{I}

One of the most interesting and, at the moment, controversial
problems in non-equilibrium statistical mechanics
is to understand the mircoscopic origins of the entropy
production in non-equilibrium processes, especially the entropy
production envisaged by irreversible thermodynamics for the usual
hydrodynamic processes. By this one means that one should start
from some totally microscopic description of a system in some intitial
state, show that in the course of time the system approaches
a stationary state, in some sense, and then calculate the entropy production
associated with this process. There is no clear definition of the
microscopic entropy production, however, and this in itself presents a
problem. One choice of an entropy was provided by Gibbs with a
definition based upon the full phase space distribution function of a
classical system. However, the
Gibbs entropy defined with respect to the phase space distribution
function remains constant in time, if the time dependence of the
distribution function is determined by the Liouville equation, as it
is for conservative, Hamiltonian systems. One solution to this
particular problem has long been discussed, the use of the Gibbs
entropy as a measure of a non-equilibrium entropy requires that every
possible trajectory in phase space (except for a set of measure zero)
be followed in time with infinite precision, but if one relaxes this
requirement and only follows trajectories to within some specified
precision, i.e. ``coarse grains" the description, then one obtains an
entropy function that increases with time. Of course, an increase with
time does not automatically imply an agreement of the entropy
production with the laws of irreversible thermodynamics. 

As a consequence of these considerations, a number of issues remain to
be resolved:
\begin{enumerate}
\item Is a coarse grained Gibbs entropy the best candidate for a
definition of a non-equilibrium entropy~? 
\item If so, how does one correctly define the coarse graining
process~? To what extent are results so obtained independent of the
coarse graining procedure~?
\item Do any definitions of entropy production lead to the laws of
irreversible thermodynamics, and if so, when and why~?
\end{enumerate}

There is, of course an enormous literature on all of these
questions. Here we wish only to discuss some recent progress in
answering them based upon the approach to a theory of irreversible
processes through dynamical systems theory. Gaspard \cite{gaspard92} has 
considered a
microscopic model of diffusion of particles in one dimension, called a
multi-baker model, where the
dynamics is modeled by a baker's transformation taking place on a 
one-dimensional lattice, where a unit square is associated to every site.
%in a long thin two-dimensional strip. Here the 
%long direction, the $x$-direction, 
%taken to
%correspond the the direction of diffusion, while the thin direction
%corresponds to all other variables that are irrelevant to the
%diffusion process and, in themselves, have no physical meaning, other
%than insuring the measure-preserving nature and reversibility of the 
%dynamics. Gaspard
%considered a finite length chain, $1 \leq x \leq L$, broken up into
%intervals of unit length and defined the dynamics on the unit square
%so that particles would be sent either to the adjacent right or left
%intervals depending upon the $x$-component of their location in the
%unit square. Tasaki and Gaspard  \cite{tasaki95} considered the case 
%where a steady gradient in particle density was maintained along the chain, 
%and were able to show that fractal-like structures formed by regions of 
%differing microscopic densities, appear in the two dimensional phase space. 
%The fractal like structures become real fractals in the thermodynamic
%limit, as $L \rightarrow \infty$. 
Here the one-dimensional lattice is identified as the configuration
space of a random walker where diffusion takes place. The baker's 
transformation exchanges points of the unit squares between neighboring 
cells and allows for a deterministic description of the random walk, i.~e. 
keeping the dynamics on the configuration space unchanged. The variables on
the unit square are irrelevant to the diffusion process and, in themselves, 
have no physical meaning, other than insuring the measure-preserving nature 
and reversibility of the dynamics. Gaspard considered a finite length 
chain, $1 \leq n \leq L,$ and defined the dynamics on the unit squares
so that particles would be sent either to the adjacent right or left
intervals depending upon their location along the expanding direction 
in the unit square. Tasaki and 
Gaspard  \cite{tasaki95} considered the case where a steady gradient in 
particle density was maintained along the chain, and were able to show that 
fractal-like structures formed by regions of differing microscopic 
densities, appear in the two dimensional phase space. 
The fractal like structures become real fractals in the infinite volume
limit, as $L \rightarrow \infty$. 

These fractal-like structures are strict consequences of the dynamics
given the presence of a density gradient produced by particle reservoirs at the
boundaries. Their importance for the theory of entropy production, as
pointed out by Gaspard \cite{gaspard97,gaspard98}, lies in the fact that 
they provide a
fundamental reason for coarse graining the distribution
function. That is, for large systems, the microscopic variations in density
take place on such fine scales that no reasonable measurement process
would or should be able to detect these variations. Gaspard 
\cite{gaspard97,gaspard98} showed
that the steady state production of entropy in this model is
in agreement with the predictions of irreversible thermodynamics and
further, that the entropy production is independent of the coarse
graining size over a wide range of possible coarse graining sizes. 
While it is not entirely clear why this procedure leads to results in
agreement with irreversible thermodynamics, the model and procedure are
sufficiently interesting and stimulating that one would hope to find
further examples so as to gain some deeper insights into the nature of
entropy production, hopefully in general, and certainly in this group
of baker transformation like models.

A closely related, independent approach to the problem of entropy
production in non-equilibrium steady states is provided by T\'el, Vollmer,
and Breymann (TVB) in a series of papers 
\cite{breymann96,vollmer97,breymann98,vollmer98}, also devoted to diffusion in
multi-baker models. These authors considered the entropy production in
measure preserving maps as well as in dissipative maps that do not preserve 
the Lebesgue measure and model systems with Gaussian thermostats. The TVB
systems also show that a coarse grained distribution function leads to
a positive entropy production in agreement with non-equilibrium
thermodynamics. In addition to considering a wider class of models
than Gaspard, they used a different coarse graining scheme that was
not devised to expose the underlying fractal structures of the SRB
measures of the two dimensional phase space associated with their
models. They also argued that their results for the entropy
production should be largely
independent of the type of coarse graining scheme used.

A generalization of the methods of Gaspard \cite{gaspard97,gaspard98}
and TVB \cite{breymann96,vollmer97,breymann98,vollmer98} was proposed
by the present authors in a recent paper \cite{gilbert99b}, where it 
was shown that a coarse-grained
form of the Gibbs entropy, which can be expressed in terms of the measures
and volumes of the coarse graining sets partitioning the phase, leads 
to an entropy production formula similar to Gaspard's and applicable to more 
general volume-preserving as well as dissipative models, such as those
considered by TVB, as well as multi-baker maps with energy flow considered
by Tasaki and Gaspard \cite{tasaki99}.

We should also mention that multi-baker models and the use of
coarse gaining methods for calculating their entropy production have
been criticized by Rondoni and Cohen\cite{rondoni99}. While some of their 
points are
indisputably correct, their approach does not suggest a better way to
proceed. So we continue this line of research in the expectation that
it will lead to some further insights into entropy production in more
realistic models, despite the shortcomings of the simplified models we
treat here.

In this paper, we extend the ideas mentioned above, particularly those
of Gaspard \cite{gaspard97,gaspard98} and Gilbert and Dorfman 
\cite{gilbert99b}, to a somewhat more complex model 
than those considered
above. We consider a deterministic version of a persistent random walk
in one dimension. The persistent random walk (PRW) is similar to the usual
random walk ,
%modeled by deterministic multi-baker maps 
but in this case
the moving particle has both a position and velocity as it moves along
a one dimensional lattice. At each lattice site the particle
encounters a scatterer which, with probability $p$ allows the particle
to continue in the direction of its velocity and with probability $q =
1-p$ reverses the direction of the velocity. This model is a limiting
version of a Lorentz lattice gas described in detail by van Velzen and
Ernst \cite{vanvelzen90}, where scatterers are distributed at random 
along the lattice sites with
some overall density per site. In the persistent random walk, the site
density is unity. Although this is clearly a random process, it can be
turned into a deterministic one by a method described by 
Dorfman, Ernst, and Jacobs \cite{dorfman95b}, based upon the baker's 
transformation, whereby new variables are added to the system such
that the dynamics in terms of these new variables allows us to replace
the stochastic scattering mechanism by a deterministic one. 
A similar mechanism has also been considered by Goldstein, Lanford and 
Lebowitz in a different context \cite{goldstein73}. Further, we can 
place many particles 
on the lattice
as long as they do not interact with each other and if we take each
individual scattering event to be independent of any others taking
place at the same instant of time. 

Here we will consider the PRW in one dimension as a model for
diffusion and entropy production, and we will describe the entropy
production in terms of the additional phase space variables needed to
make the system deterministic. That an irreversible entropy production
is to be associated with this process follows from the fact that for
systems with periodic boundary condiditons, any
intitial variations in the probability of finding a particle at given 
positions
will eventually vanish and the probability will become uniform with
time. Thus information is lost about the position of the particle and
the entropy of the system is thereby increased. It is of some interest
to see how this loss of information is reflected in the phase space
distribution function, and how the entropy so produced is related to
the entropy which is the subject of irreversible thermodynamics. We
will see that the analysis of the PRW model has some features that one
hopes would be more general, such as the clear independence of the
entropy production on the sizes of the coarse graining regions, provided
these regions are not too small, and the coincidence of the
production of the coarse grained Gibbs entropy with the entropy
production of irreversible thermodynamics, previously noted by Gaspard,
TVB and the present authors in multi-baker maps.

The organization of this paper is as follows. Section~\ref{BIS}
gives a simple description of the phenomenological approach to 
entropy production in a simple one-dimensional diffusive system.
In Sec.~\ref{II}, we describe in more detail the PRW process. In
Sec.~\ref{III} we describe a two-dimensional, measure preserving map 
that reproduces the PRW on a macroscopic scale and discuss its time-reversal
properties. We consider a system of non-interacting
particles with a steady density gradient, produced by particle
reservoirs at each end of the system. We then obtain the steady state
invariant measure using techniques due to Gaspard and Tasaki 
\cite{tasaki95}. In
Sec.~\ref{IV}  we develop a symbolic dynamics for the diffusion process,
as it is refected in phase space, and, in Sec.~\ref{V}, we apply this
symbolic dynamics to compute the rate of entropy production. We
conclude with remarks and a discussion of open questions in Sec.~\ref{VI}.

\section{\label{bis}Phenomenological approach}
%\eqlabel{bis}

Before turning to the description of the persistent random walk in
Sec. \ref{II}, we briefly discuss the phenomenological approach to
entropy production in a one-dimensional system.

Let us consider a stochastic diffusive process that is driven away from
equilibrium. More specifically, we have in mind a
system where a gradient of density is imposed by appropriate boundary
conditions. Let the system be essentially one-dimensional by imposing
translational invariance in the other spatial directions. The relevant spatial
direction is denoted by $x,$ $0 < x < L + 1.$ At the boundaries,
$x = 0$ and $x = L + 1,$  we put particle reservoirs with 
respective particle densities $w_-$ and $w_+.$ The reservoirs keep
on feeding the system with new particles and those particles that 
exit never come back.

The probability density of a tracer particle is $w_t(x)$ and obeys
the mass conservation law
\begin{equation}
\fr{\partial w_t(x)}{\partial t} + \vec{\nabla}\cdot\vec{j}_t(x) = 0,
\label{OVmasscons}
\end{equation}
%\eqlabel{OVmasscons}
where $\vec{j}_t(x)$ is the associated current.

To follow the phenomenological approach of non-equilibrium thermodynamics,
we have to supplement the mass conservation law with a linear law that
relates the current $\vec{j}_t(x)$ to the gradient of the density 
$\vec{\nabla}w_t(x),$ 
\begin{equation}
\vec{j}_t(x) = -D \vec{\nabla}w_t(x),
\label{OVfickslaw}
\end{equation}
%\eqlabel{OVfickslaw}
where $D$ is the diffusion coefficient associated with that process.
This is known as Fick's law \cite{degroot84}.

Combining Eqs.~(\ref{OVmasscons}-\ref{OVfickslaw}) and assuming that the
diffusion coefficient is constant, we obtain the
Fokker-Planck equation for diffusion,
\begin{equation}
\fr{\partial w_t(x)}{\partial t} = D \nabla^2 w_t(x).
\label{OVfokkerplanck}
\end{equation}
%\eqlabel{OVfokkerplanck}
The stationary solution of this equation is found by 
imposing the boundary conditions
$w_t(0) = w_-$ and $w_t(L + 1) = w_+,$
\begin{equation}
w(x) = w_- + (w_+ - w_-)\fr{x}{L + 1}.
\label{OVwss}
\end{equation}
%\eqlabel{OVwss}

The connection with the second law of thermodynamics is made by 
considering the entropy whose local density is defined as
\begin{equation}
s_t(x) = -\{\log[w_t(x)] - 1\},
\label{OVlocent}
\end{equation}
%\eqlabel{OVlocent}
and $S(t) = \int_0^{L + 1}dx w_t(x) s_t(x)$ is the macroscopic entropy. 
Taking the 
derivative of the integrand with respect to time, we find successively
\begin{eqnarray}
\fr{\partial w_t(x)s_t(x)}{\partial t} &=& -\fr{\partial w_t(x)}{\partial t}
\log[w_t(x)]\nonumber\\
&=&[\vec{\nabla}\cdot\vec{j}_t(x)]\log[w_t(x)]\nonumber\\
&=&\vec{\nabla}\cdot\{\vec{j}_t(x)\log[w_t(x)]\} + \fr{\vec{j}_t(x)^2}{D w_t(x)},
\label{OVentch1}
\end{eqnarray}
%\eqlabel{OVentch1}
where we used Eq.~(\ref{OVmasscons}) in the second line and Eq.~(\ref{OVfickslaw})
in the third one. Equation (\ref{OVentch1}) has the form of a local entropy
balance \cite{degroot84}
\begin{equation}
\fr{\partial w_t(x)s_t(x)}{\partial t}= 
- \vec{\nabla}\cdot \vec{J}_s^{\rm tot}(x,t) + \sigma_t(x),
\label{OVentbal1}
\end{equation}
%\eqlabel{OVentbal1}
where $\vec{J}_s^{\rm tot}(x,t)$ is the total entropy flow and 
$\sigma_t(x) > 0$
is the entropy source term. Equation (\ref{OVentbal1}) can be rewritten
in a slightly different form
\begin{equation}
w_t(x)\fr{ds_t(x)}{dt}= 
- \vec{\nabla}\cdot \vec{J}_s(x,t) + \sigma_t(x),
\label{OVentbal2}
\end{equation}
%\eqlabel{OVentbal2}
where the entropy flux is the difference
\begin{equation}
\vec{J}_s(x, t) = \vec{J}_s^{\rm tot}(x,t) - w_t(x)s_t(x)\vec{j}_t(x),
\label{OVentflux}
\end{equation}
%\eqlabel{OVentflux}
between the total entropy flux and a convective term.

Using Eq.~(\ref{OVfickslaw}) again, we find that the local rate of entropy 
production 
$\sigma_t(x)$ in Eq.~(\ref{OVentch1}) can be rewritten in terms of the 
density $w_t(x)$ only~:
\begin{equation}
\sigma_t(x) = D \fr{[\vec{\nabla}w_t(x)]^2}{w_t(x)}.
\label{OVtdepentprod}
\end{equation}
%\eqlabel{OVtdepentprod}
In the stationary state, Eq.~(\ref{OVwss}), the local rate of entropy production
becomes
\begin{equation}
\sigma(x) = D\fr{(w_+ - w_-)^2}{(L + 1)^2 w(x)}.
\label{OVssentprod}
\end{equation}
%\eqlabel{OVssentprod}

\section{\label{II}Persistent Random Walk -- Definitions}
%\eqlabel{II}

We consider the following process~: a particle on a 
one-dimensional lattice moves from site to site with a given velocity
$\pm 1$ and, at each time step, is scattered forward with probability 
$p$ and backwards with probability $q,$ such that $p + q = 1$. Thus a 
particle located at site $n\in{\bf Z}$ with velocity $v = \pm 1$ will move 
to $n \pm 1$ conserving or reversing its velocity with the probabilities~:
\begin{equation}\label{PRWdef}
(n, v)\quad\rightarrow\quad
\left\{
\begin{array}{lr}
(n + v, v)\quad &{\rm with\:probability\:}p,\\
(n - v, - v)\quad &{\rm with\:probability\:}q.
\end{array}
\right.
\end{equation}
%\eqlabel{PRWdef}

The diffusion coefficient for this process is 
\cite{vanbeijeren82,vanvelzen90,vanbeijeren93}
\begin{equation}\label{PRWDifcoef}
D = \fr{p}{2q}.
\end{equation}
%\eqlabel{PRWDifcoef}

\begin{figure}[htb]
\centerline{\psfig{figure=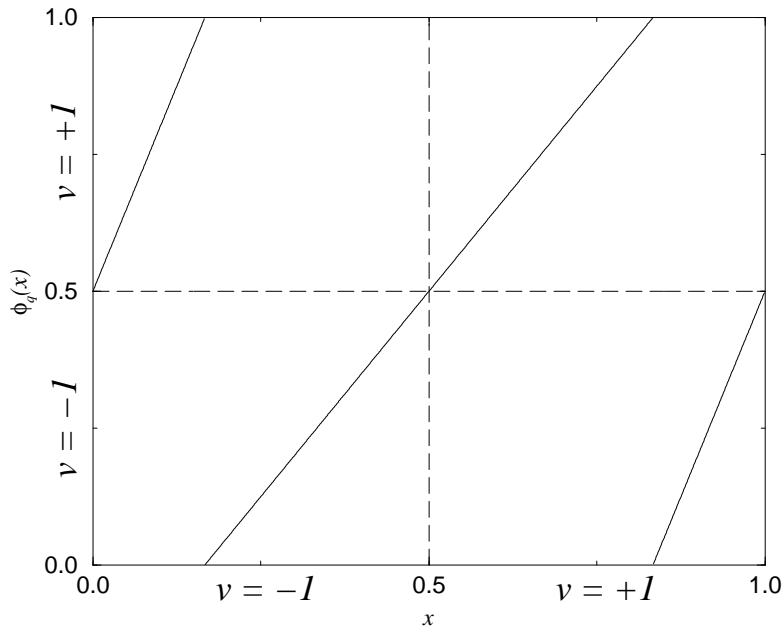,width=12cm}}
\caption{\label{figPRWvmap}{\footnotesize The map $\phi_q$ defined by 
Eq. (\ref{PRWvmap})
mimics the probability rules, Eq. (\ref{PRWdef}), for the velocity vector
$v$.}}
\end{figure}

\begin{figure}[htb]
\centerline{\psfig{figure=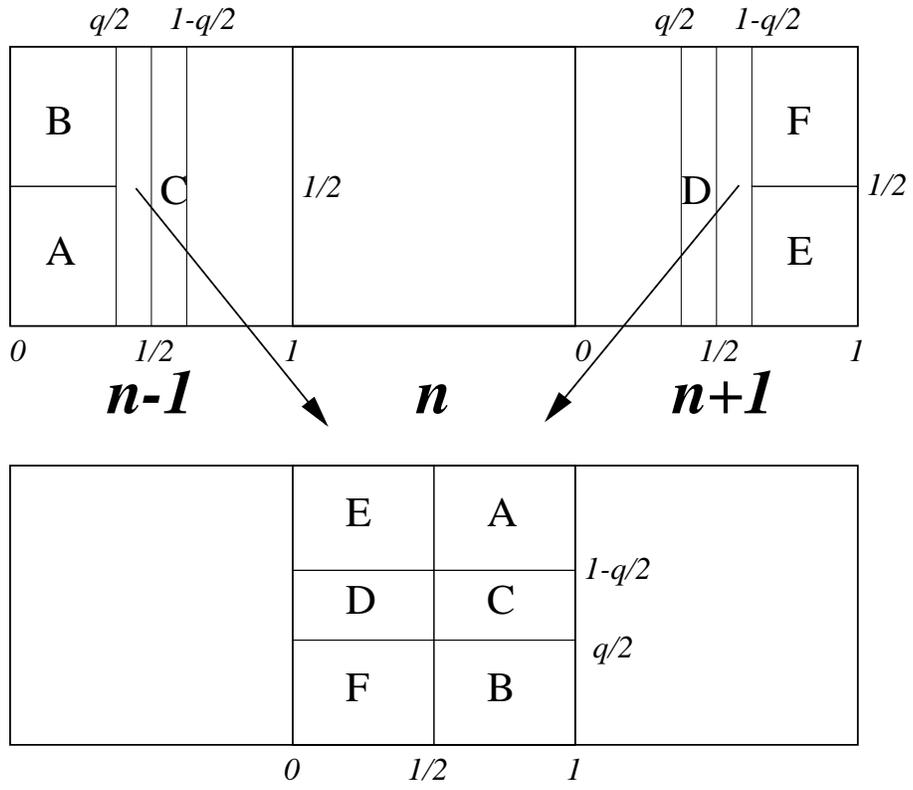,width=12cm}}
\caption{\label{figPRWmap}{\footnotesize The multi-baker map, 
Eq. (\ref{PRWmap}), that models the persistent random walk, 
Eq. (\ref{PRWdef}).}}
\end{figure}

\begin{figure}[htb]
\centerline{\psfig{figure=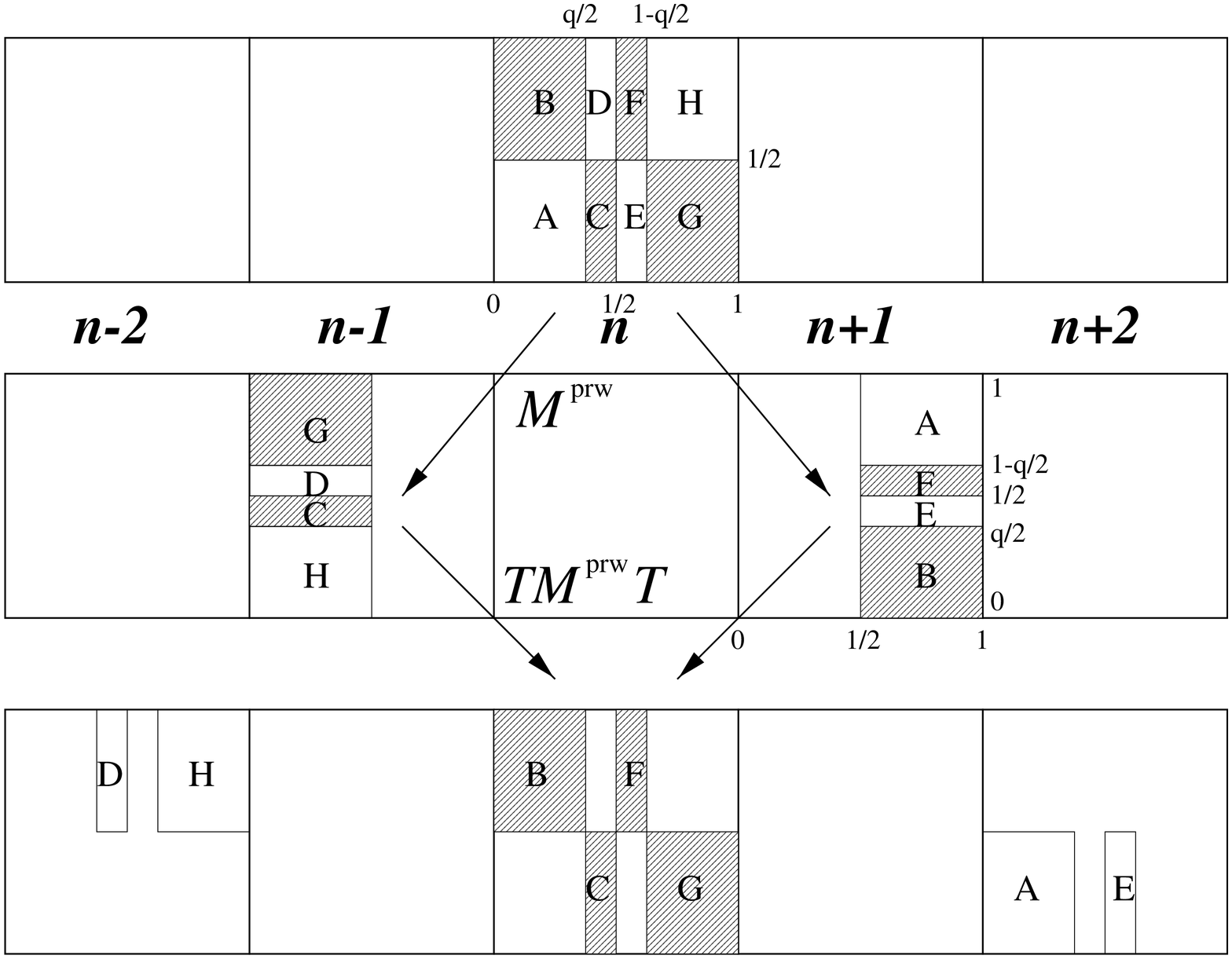,width=12cm}}
\caption{\label{figPRWTMTM}{\footnotesize The composition of $\Mprw$ with 
$T,$ as in Eq. (\ref{PRWTMTM}).}}
\end{figure}

\begin{figure}[htb]
\centerline{\psfig{figure=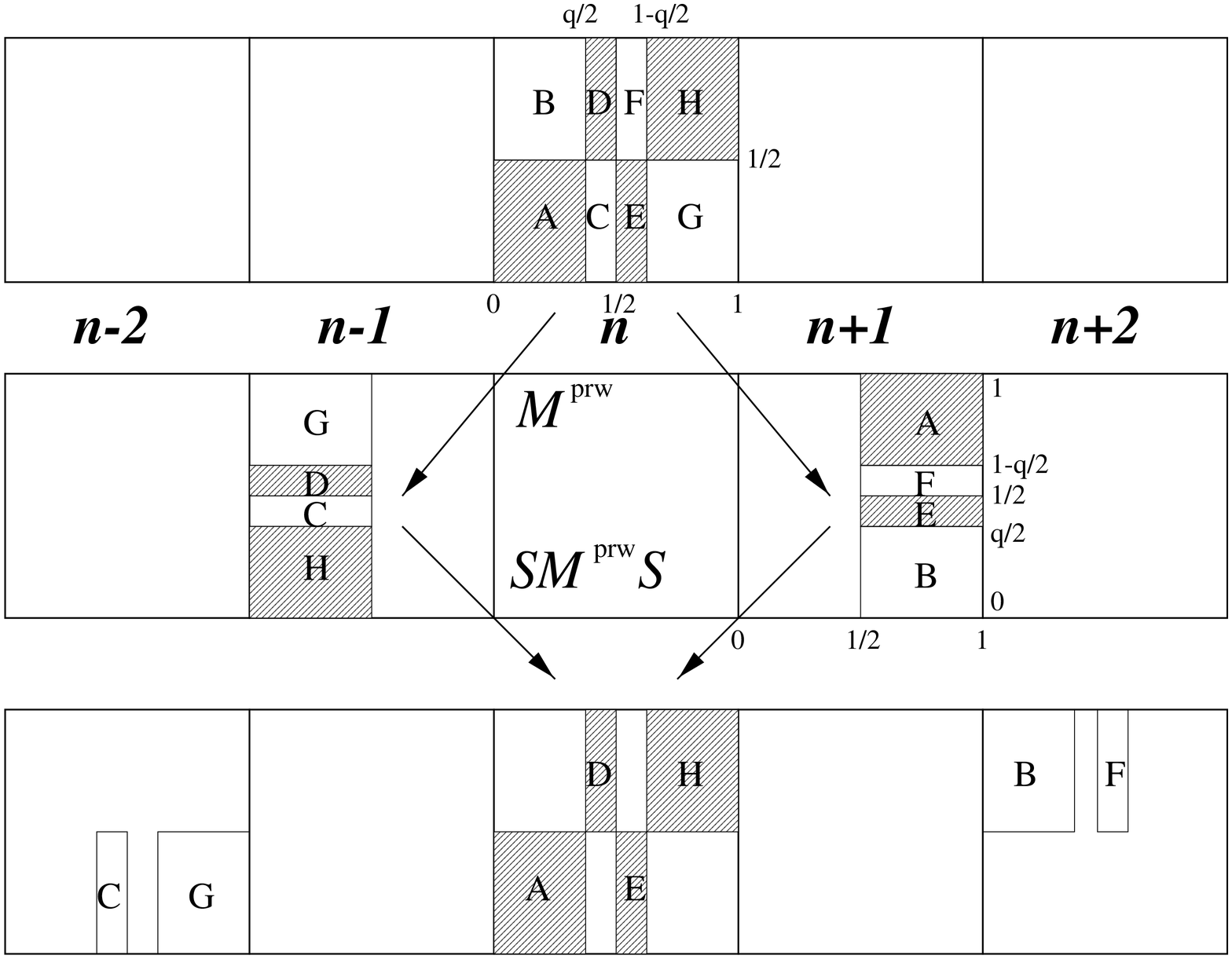,width=12cm}}
\caption{\label{figPRWSMSM}{\footnotesize The composition of $\Mprw$ with 
$S,$ similar to Eq. (\ref{PRWTMTM}).}}
\end{figure}

\begin{figure}[htb]
\centerline{\psfig{figure=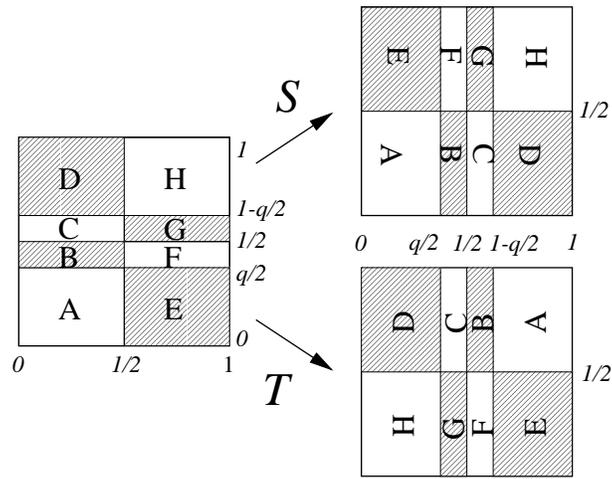,width=8cm}}
\caption{\label{figPRWTS}{\footnotesize The action of $S$ and $T$ on the 
elements of the partition induced by $\Mprw.$}}
\end{figure}

To construct a deterministic model of this process, we start with a procedure 
similar to the one used by Dorfman, Ernst and Jacobs \cite{ernst95,dorfman95b},
and consider 
the velocity of the particle only and associate to it a unit interval.
Let us divide the interval into two halves, corresponding to the two possible
values of the velocity. For definiteness, let us assign the first half to 
$v = -1$ and the second to $v = + 1.$ According to Eq. (\ref{PRWdef}), a 
particle has a probability $p$ of keeping its velocity unchanged and $q$
to reverse it. Therefore we have to further subdivide both halves into two 
parts of respective lengths $p/2$ and $q/2$, the first of which must be 
mapped onto the same half and the second onto the other half. Each of these
branches must be linear and onto in order to model the independent 
process of velocity flips. This is illustrated in Fig. 
(\ref{figPRWvmap}) for the case $q = 1/3.$
Therefore a one-dimensional map whose dynamics
mimics the random sequence of the velocities is given by
\begin{equation}
\phi_q(x) = 
\left\{
\begin{array}{l@{\quad}l}
\fr{x}{q} + \fr{1}{2},& 0\leq x < \fr{q}{2},\\
\fr{x - \fr{q}{2}}{p},& \fr{q}{2} \leq x < 1 - \fr{q}{2},\\
\fr{x - 1 + \fr{q}{2}}{q},&1 - \fr{q}{2} \leq x < 1.
\end{array}
\right.
\label{PRWvmap}
\end{equation}
%\eqlabel{PRWvmap}
The generalization of $\Phi_q(x)$ to a multi-baker map is straightforward~:
\begin{equation}\label{PRWmap}
\Mprw (n, x, y) =
\left\{
\begin{array}{lr}
\left(n + 1, \fr{x}{q} + \fr{1}{2}, q\left(y - \fr{1}{2}\right) + 1\right), 
&0 \leq x < \fr{q}{2},\:0 \leq y < \fr{1}{2},\\
\left(n + 1, \fr{x}{q} + \fr{1}{2}, q\left(y - \fr{1}{2}\right)\right), 
&0 \leq x < \fr{q}{2},\:\fr{1}{2}\leq y < 1,\\
\left(n - 1, \fr{x - q/2}{p}, py + \fr{q}{2}\right), 
&\fr{q}{2}\leq x < \fr{1}{2},\\
\left(n + 1, \fr{x - q/2}{p}, py + \fr{q}{2}\right), 
&\fr{1}{2}\leq x < 1 - \fr{q}{2},\\
\left(n - 1, \fr{x - 1 + q/2}{q}, q\left(y - \fr{1}{2}\right) + 1\right), 
&1 - \fr{q}{2} \leq x < 1,
\:0 \leq y < \fr{1}{2},\\
\left(n - 1, \fr{x - 1 + q/2}{q}, q\left(y - \fr{1}{2}\right)\right), 
&1 - \fr{q}{2} \leq x < 1,\:\fr{1}{2}\leq y < 1.\\
\end{array}
\right.
\end{equation}
%\eqlabel{PRWmap}
This map is shown in Fig(\ref{figPRWmap}).
We note that if one restricts this map to a single unit square, by ignoring
the changes in the site index, $n$, this map is reversing 
under the time-reversal operators of the usual baker map 
$T(x,y) = (1 - y, 1 - x)$ and $S(x,y) = (y,x),$ i.~e.
\begin{eqnarray}
T \circ \Mprw \circ T (x, y) &=& \Mprw^{-1} (x, y),\label{trT}\\
S \circ \Mprw \circ S (x, y) &=& \Mprw^{-1} (x, y),\label{trS}.
\end{eqnarray}
However, the full multi-baker map, with the changes in the site index
taken into account, reversing under either of these two time-reversal operators. Indeed, one finds that, 
depending
on $x$ and $y,$ the composition 
\begin{equation}
T \circ \Mprw \circ T \circ \Mprw (n, x, y) = 
\left(\left\{
\begin{array}{l}
n + 2\\
n\\
n - 2
\end{array}
\right\}
,x, y\right),
\label{PRWTMTM}
\end{equation}
%\eqlabel{PRWTMTM}
and similarly for $S.$ This is displayed in Figs. 
(\ref{figPRWTMTM}-\ref{figPRWSMSM}).
It is interesting to note that we can combine the action of $S$ and $T$
to form a hybrid operator $U$ that is reversing. Indeed, as seen in 
Fig. (\ref{figPRWTS}), 
$S$ and $T$ act on the square in such a way that
the images of the elements of the natural partition of the square under
these two operators do not overlap.
Therefore $U,$ defined by
\begin{equation}
U(x, y) = 
\left\{
\begin{array}{l@{\quad}l@{\quad}l}
T(x, y),&
0\leq x < \fr{1}{2}, & \fr{q}{2}\leq y < \fr{1}{2},\\
&&1 - \fr{q}{2} \leq y < 1,\\
&\fr{1}{2} \leq x < 1, &
0 \leq y < \fr{q}{2},\\
&&\fr{1}{2}\leq y < 1 - \fr{q}{2},\\
S(x, y)&{\rm otherwise},&
\end{array}
\right.
\label{PRWU}
\end{equation}
%\eqlabel{PRWU}
is a reversing operator for $\Mprw.$ However, contrary to $T$ and $S,$ 
$U$ is not an involution,
i.e. $U\circ U(x,y) \neq (x,y),$ and therefore $U$ is not necessarily a 
reversing symmetry of any power of $\Mprw.$ Actually $U\circ U$ is an
involution,
\begin{equation}
U^4(x,y) = U\circ U\circ U\circ U(x,y) = (x, y).
\label{PRWU4}
\end{equation}
%\eqlabel{PRWU4}
This implies that the conjugation of $U$ with odd powers of $\Mprw$ yields
its inverse, while the conjugation with even powers leaves the map unchanged.
In other words, $U$ is a reversing symmetry of $\Mprw^n$ for $n$ odd but 
a symmetry for $n$ even~:
\begin{equation}
U\circ \Mprw^n \circ U = 
\left\{
\begin{array}{l@{\quad}l}
\Mprw^{-n},&n\:{\rm odd},\\
\Mprw^n,&n\:{\rm even},
\end{array}
\right.
\label{PRWUMU}
\end{equation}
%\eqlabel{PRWUMU}
This property of a reversing operator can be contrasted with the existence
of two reversing symmetries for the maps on the unit square, which itself
implies the existence of a non-trivial symmetry, i.e. the composition of $T$
and $S.$ Those properties are well established \cite{lamb98}, while the
property described by Eq. (\ref{PRWUMU}) is new, to our knowledge.

The weakened reversibility of the map $\Mprw,$ Eq. (\ref{PRWUMU}), 
still allows the map to have
%enough to ensure 
a properly behaved dynamical entropy production as we
will see in Sec. \ref{V}.

\section{\label{III}Stationary State under Flux Boundary Conditions}
%\eqlabel{III}

As with the open multi-baker map \cite{tasaki95,gaspard97,gaspard98}, 
we impose flux boundary conditions. We 
consider a chain of $L$ sites and study the distribution of an infinite
number of copies of identical systems imposing that, in average, 1
particle is present at the left-end of the chain, regardless of its velocity 
and $L + 2$ particles at the right-end.

The stationary measure is found by considering the following
cumulative functions~:
\begin{eqnarray}
\label{PRWdefcfm}
G^{(-)}(n, x, y) &=& \int_0^x dx'\int_0^y dy' \rho(n, x', y'), 
\quad 0 \leq x < \fr{1}{2},\\
\label{PRWdefcfp}
G^{(+)}(n, x, y) &=& \int_{1/2}^x dx'\int_0^y dy' \rho(n, x', y'), 
\quad \fr{1}{2} \leq x < 1, 
\end{eqnarray}
%\eqlabel{PRWdefcfm-p}
where $\rho(n, x, y)$ is the corresponding density function. Because the
map is piecewise uniformly expanding along the $x$ direction, the invariant
measure is uniform along the $x$-direction and we therefore have 
\begin{equation}\label{PRWxunifcf}
G^{(-)}(n, x, y) = 2 x g^{(-)}(n, y), \quad
G^{(+)}(n, x, y) = (2 x - 1) g^{(+)}(n, y),
\end{equation}
%\eqlabel{PRWxunifcf}
where the functions $g^{(-)}$ and $g^{(+)}$ are solutions of the following 
set of equations~:
\begin{equation}
\label{PRWeqcf}
g\m(n, y) = 
\left\{
\begin{array}{l}
qg\p\left(n \pm 1, \fr{y}{q} + \fr{1}{2}\right) - 
qg\p\left(n \pm 1, \fr{1}{2}\right), \quad 0\leq y < \fr{q}{2},\\
qg\p\left(n \pm 1, 1\right) - qg\p\left(n \pm 1, \fr{1}{2}\right)
+ pg\m\left(n \pm 1, \fr{y - q/2}{p}\right),\\
\hspace{1cm} \fr{q}{2}\leq y < 1 - \fr{q}{2},\\
qg\p\left(n \pm 1, 1\right) - qg\p\left(n \pm 1, \fr{1}{2}\right)
+ pg\m\left(n \pm 1, 1\right)\\
\phantom{+}+ qg\p\left(n \pm 1,\fr{y - 1}{q} + \fr{1}{2}\right),
\quad 1 - \fr{q}{2}\leq y < 1,
\end{array}
\right.
\end{equation}
%\eqlabel{PRWeqcf}

\begin{figure}[phtb]
\centerline{\psfig{figure=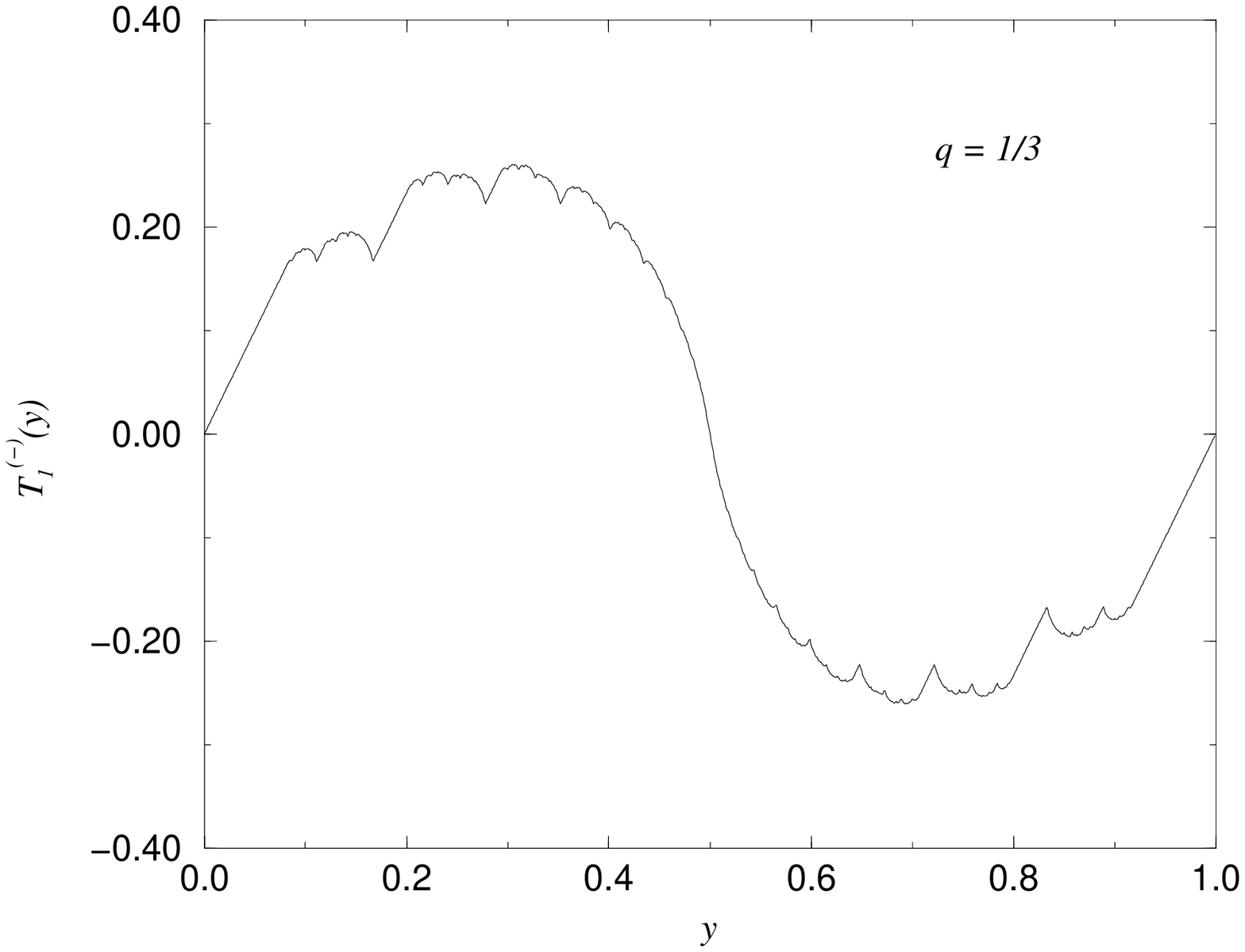,width=8cm}
\psfig{figure=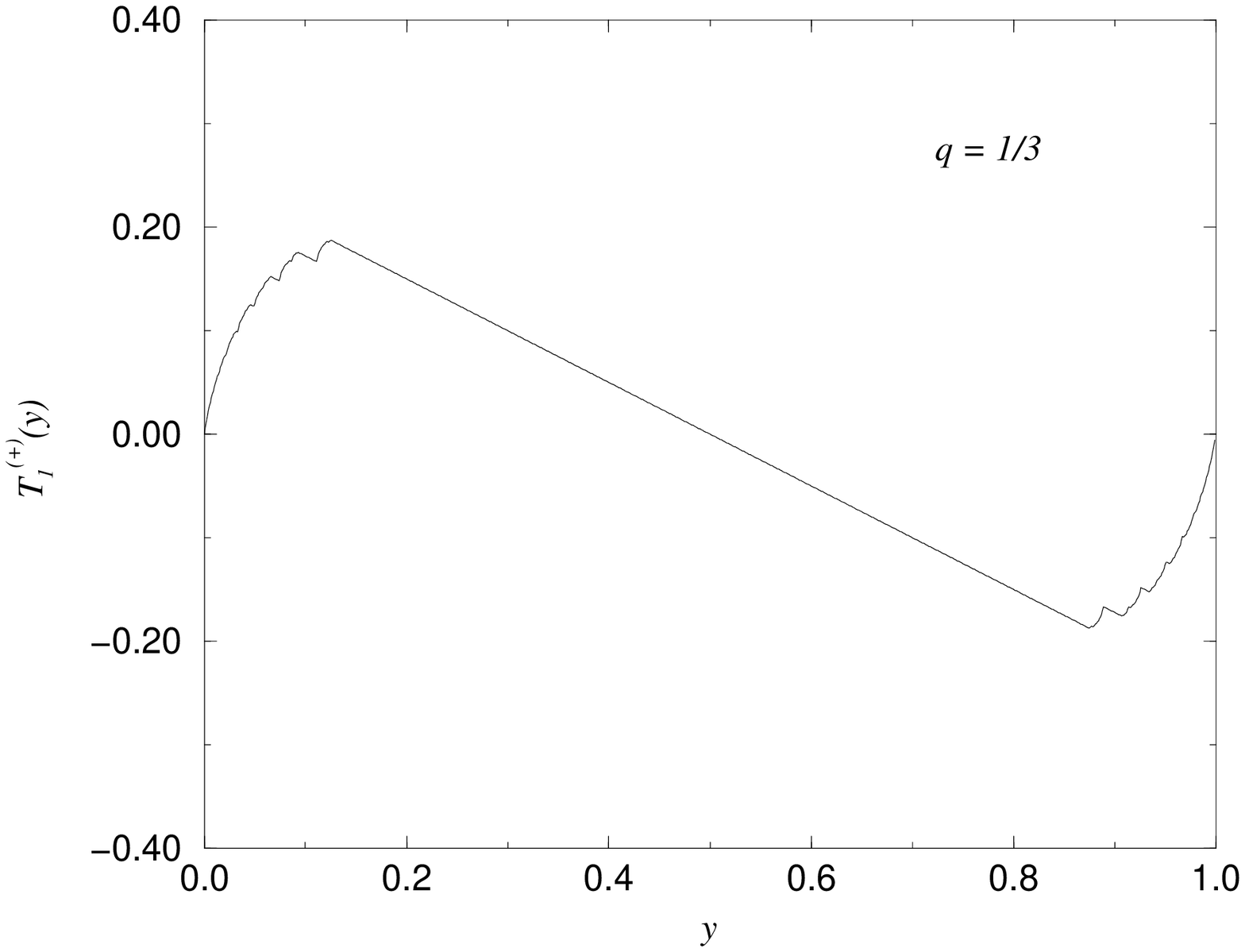,width=8cm}}
\centerline{\psfig{figure=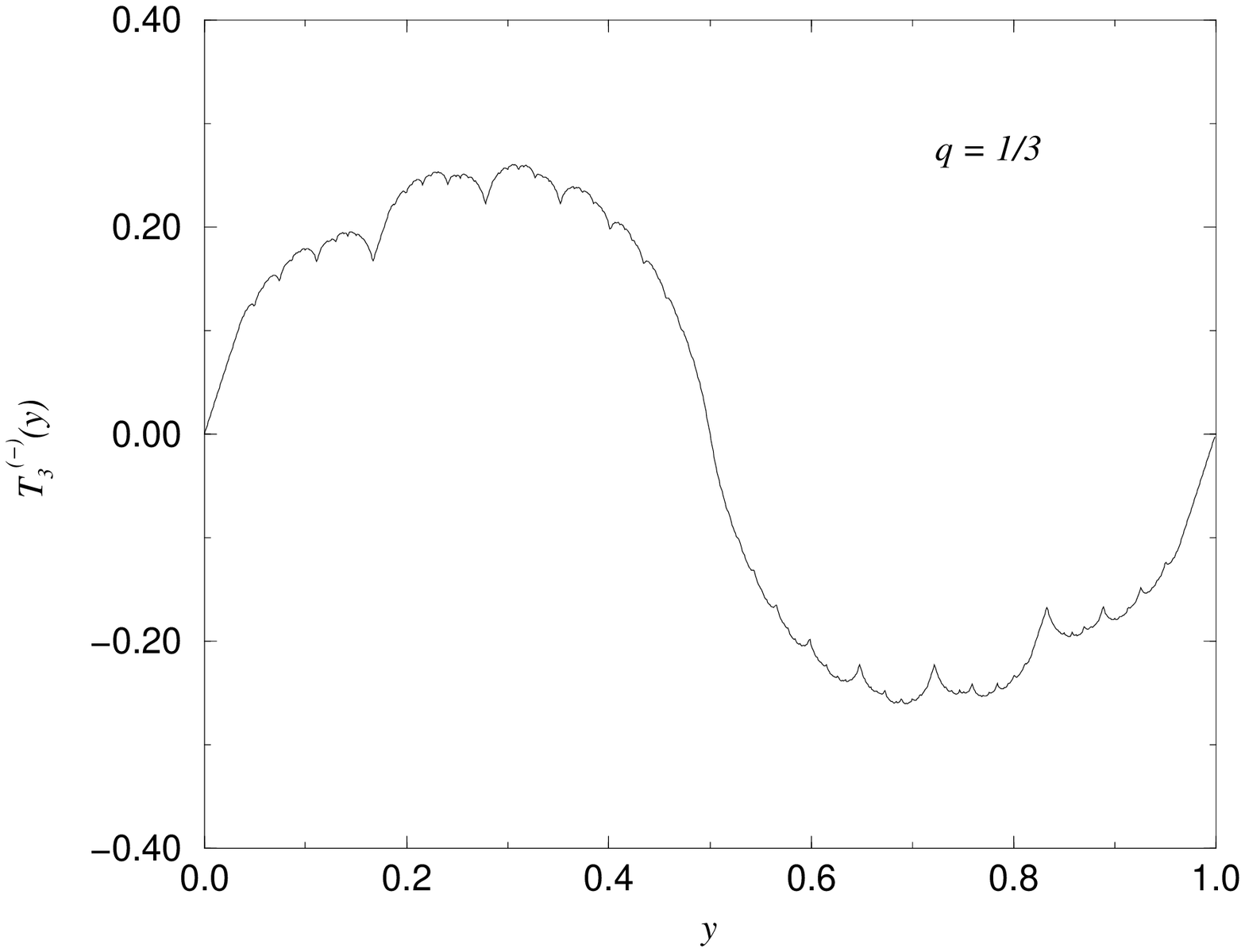,width=8cm}
\psfig{figure=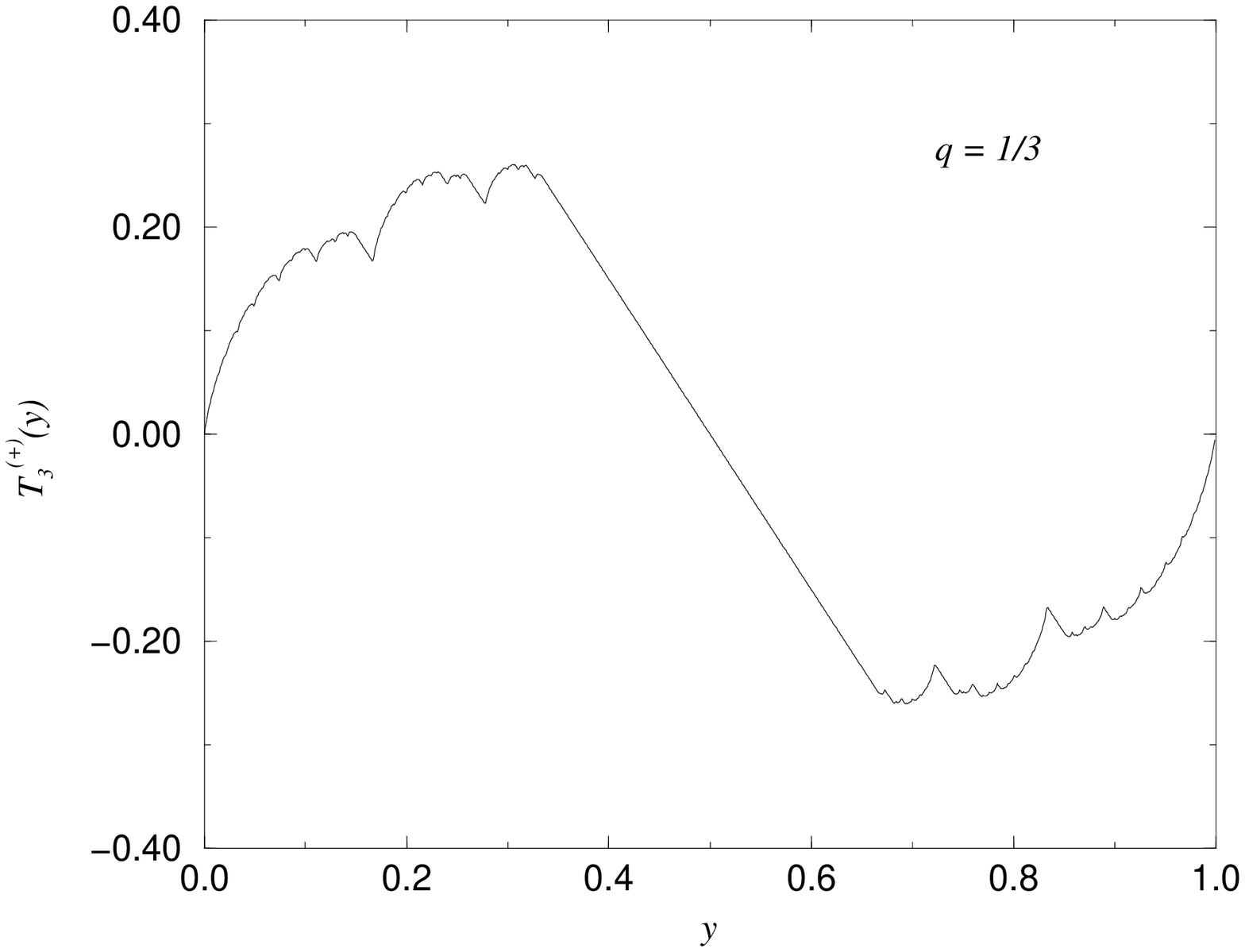,width=8cm}}
\centerline{\psfig{figure=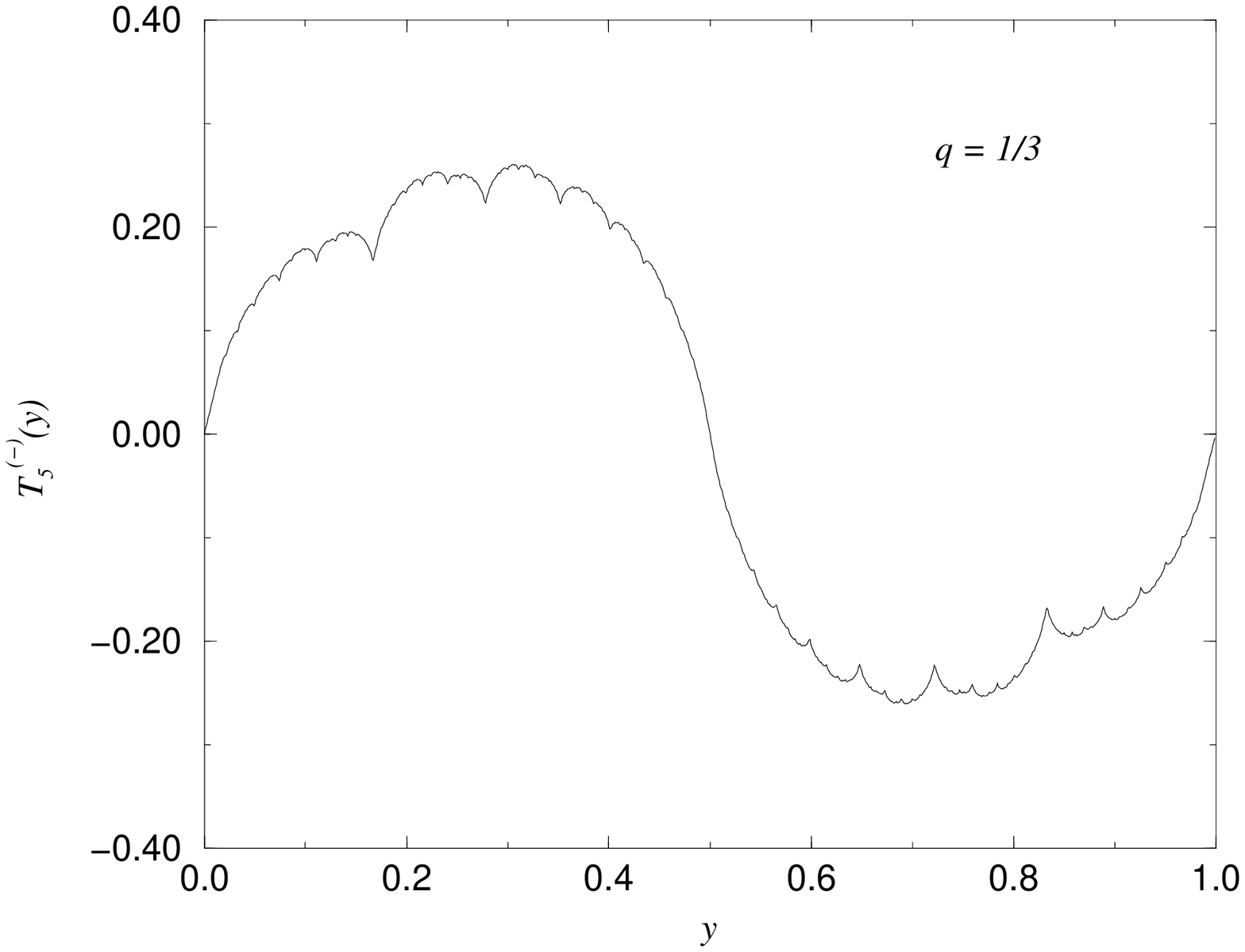,width=8cm}
\psfig{figure=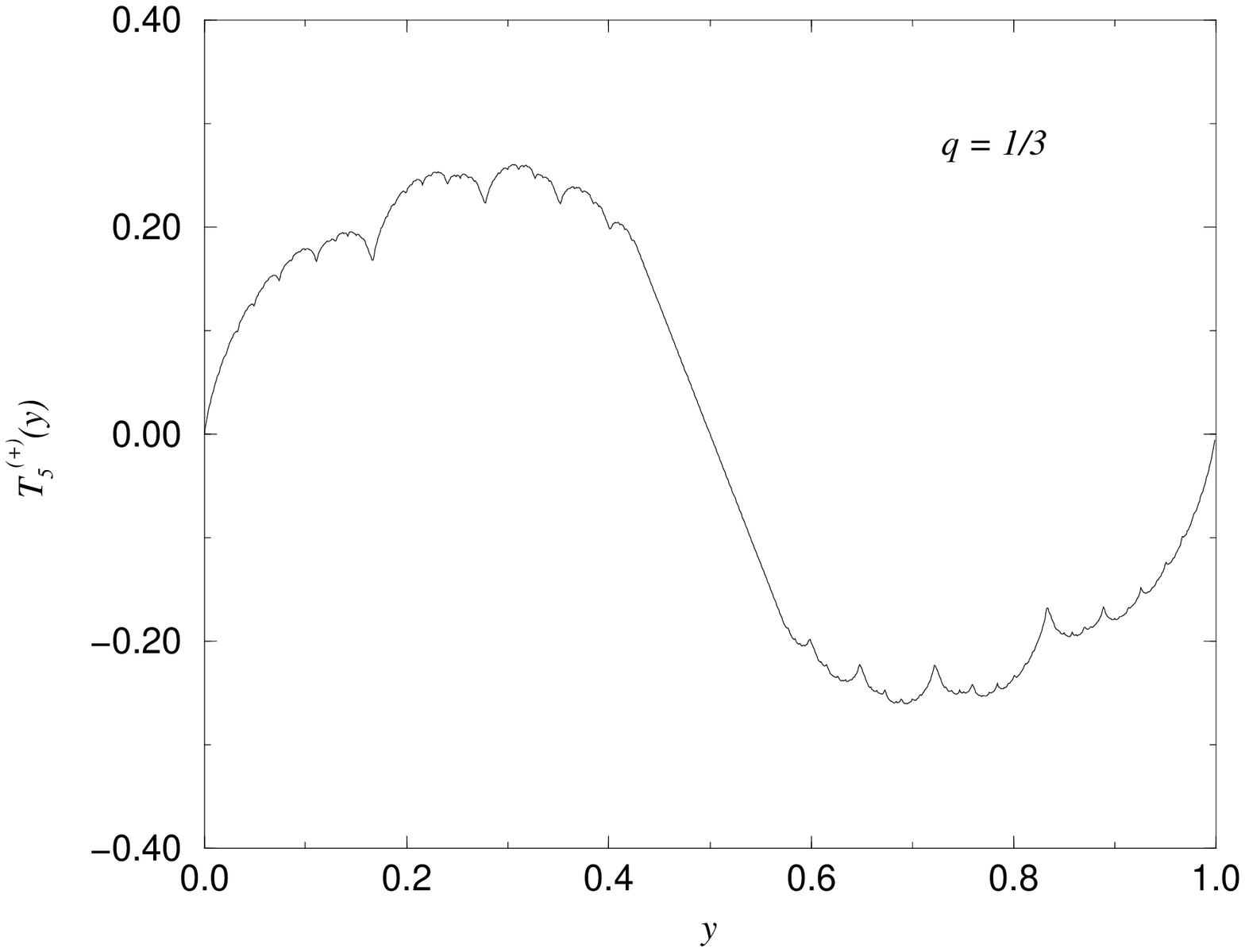,width=8cm}}
\caption{\label{figPRWincT}{\footnotesize The incomplete Takagi functions 
$T\p_n(y),$ Eq.
(\ref{PRWtpmq}),
on the right and left respectively, for $q = 1/3$ and $n = 1, 3, 5,$ from
top to bottom. The size 
of the chain is $L = 100.$}}
\end{figure}

With flux boundary conditions, 
\begin{equation}
g^{(-)}(0, y) + g^{(+)}(0, y) = y,\quad 
g^{(-)}(L + 1, y) + g^{(+)}(L + 1, y) = (L + 2)y,
\label{PRWfbc}
\end{equation}
%\eqlabel{PRWfbc}
the solutions of 
(\ref{PRWeqcf}) for $y = 1$ are readily found to be 
\begin{equation}\label{PRWy1cf}
g^{(\pm)}(n, 1) = \fr{n + 1}{2} \mp \fr{1}{4q}.
\end{equation}
%\eqlabel{PRWy1cf}
So that, for arbitrary $y$, the solution takes the form
\begin{equation}
\label{PRWsolcf}
g\p(n, y) = \left(\fr{n + 1}{2} \mp \fr{1}{4q}\right)y \pm T\p_n(y),
\end{equation}
%\eqlabel{PRWsolcf}
where the class of functions $T\p_n$ is a one-parameter family, i.e. 
implicitly dependent on the scattering parameter $q$, of generalized 
incomplete Takagi functions,
\begin{equation}\label{PRWtpmq}
T\p_n(y) = 
\left\{
\begin{array}{lr}
\fr{p}{2q} y - q T\m_{n\mp 1}\left(\fr{y}{q} + \fr{1}{2}\right), 
&0 \leq y < \fr{q}{2},\\
\fr{1/2 - y}{2} + p T\p_{n\mp 1}\left(\fr{y - q/2}{p}\right), 
&\fr{q}{2} \leq y < 1 - \fr{q}{2},\\
- \fr{p}{2q} (1 - y) - q T\m_{n\mp 1}\left(\fr{y - 1}{q} + \fr{1}{2}\right), 
&1 - \fr{q}{2} \leq y < 1,
\end{array}
\right.
\end{equation}
%\eqlabel{PRWtpmq}
where $1\leq n\leq L$ and the boundary conditions on the functions 
$T\p_n(y)$ are 
\begin{equation}\label{PRWBCTq}
T^{(+)}_0(y) = 0,\quad T^{(-)}_{L + 1}(y) = 0.
\end{equation}
%\eqlabel{PRWBCTq}
In Fig. (\ref{figPRWincT}), we display the functions
$T\p_n$ for the sites $n = 1, 3, 5$ on a chain of $L = 100$ sites and the
value of the scattering parameter $q = 1/3.$

\begin{figure}[htb]
\centerline{\psfig{figure=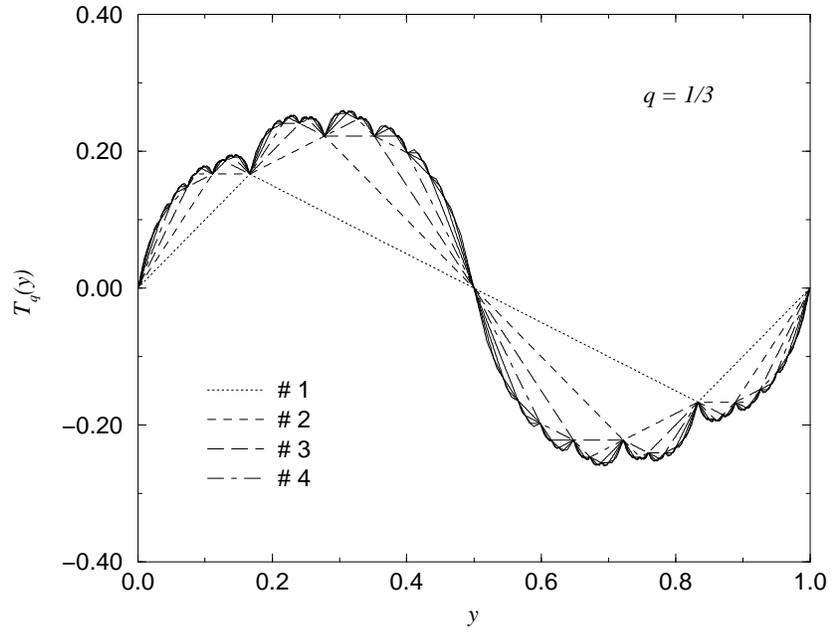,width=12cm}}
\caption{\label{figPRWTq}{\footnotesize A recursive computation of the 
generalized Takagi 
function $T_q,$ Eq. (\ref{PRWtq}) for $q = 1/3.$ The legend indicates the
numbers of the first four iterates. A total of 10 iterates are displayed.}}
\end{figure}

\begin{figure}[phtb]
\centerline{\psfig{figure=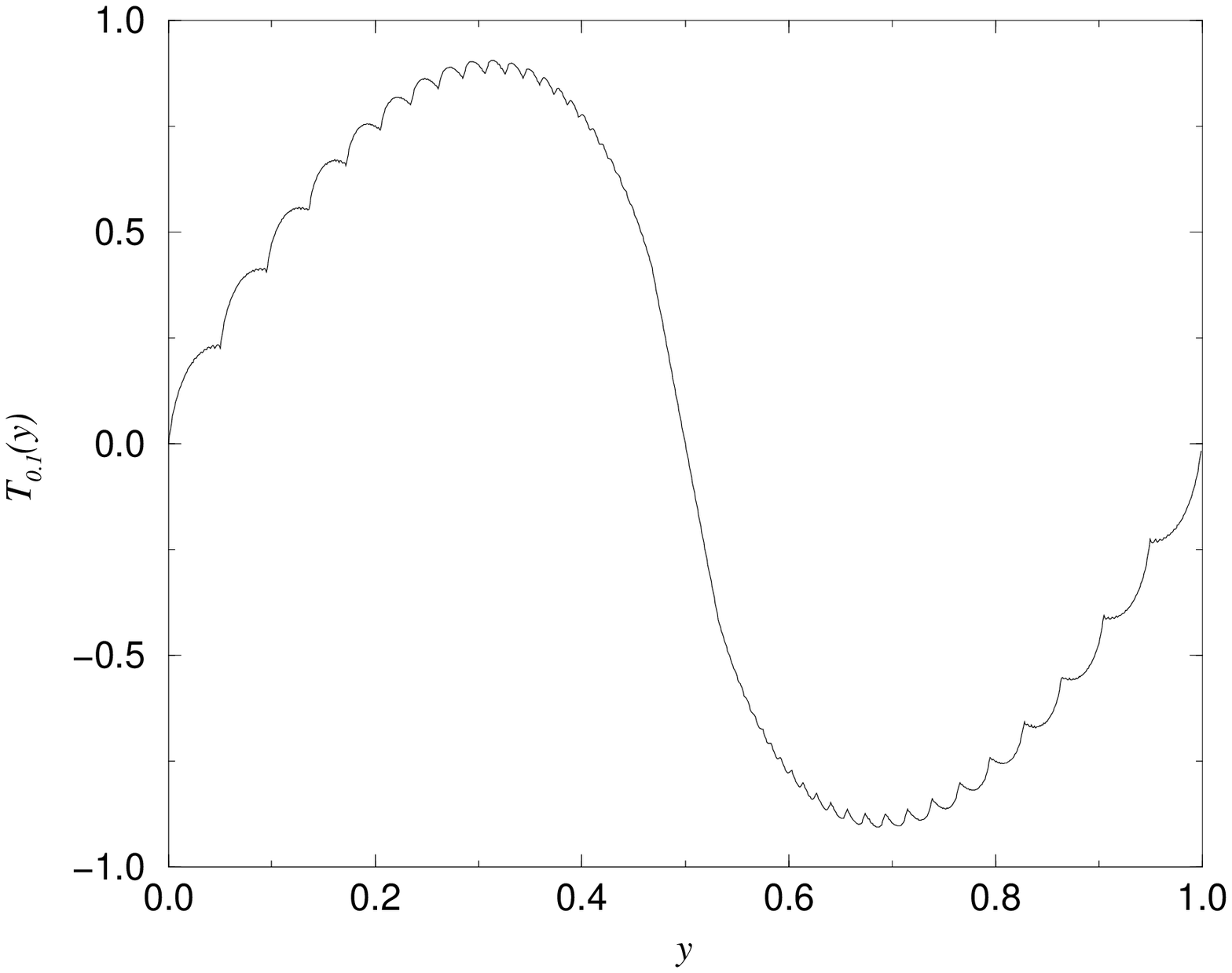,width=8cm}
\psfig{figure=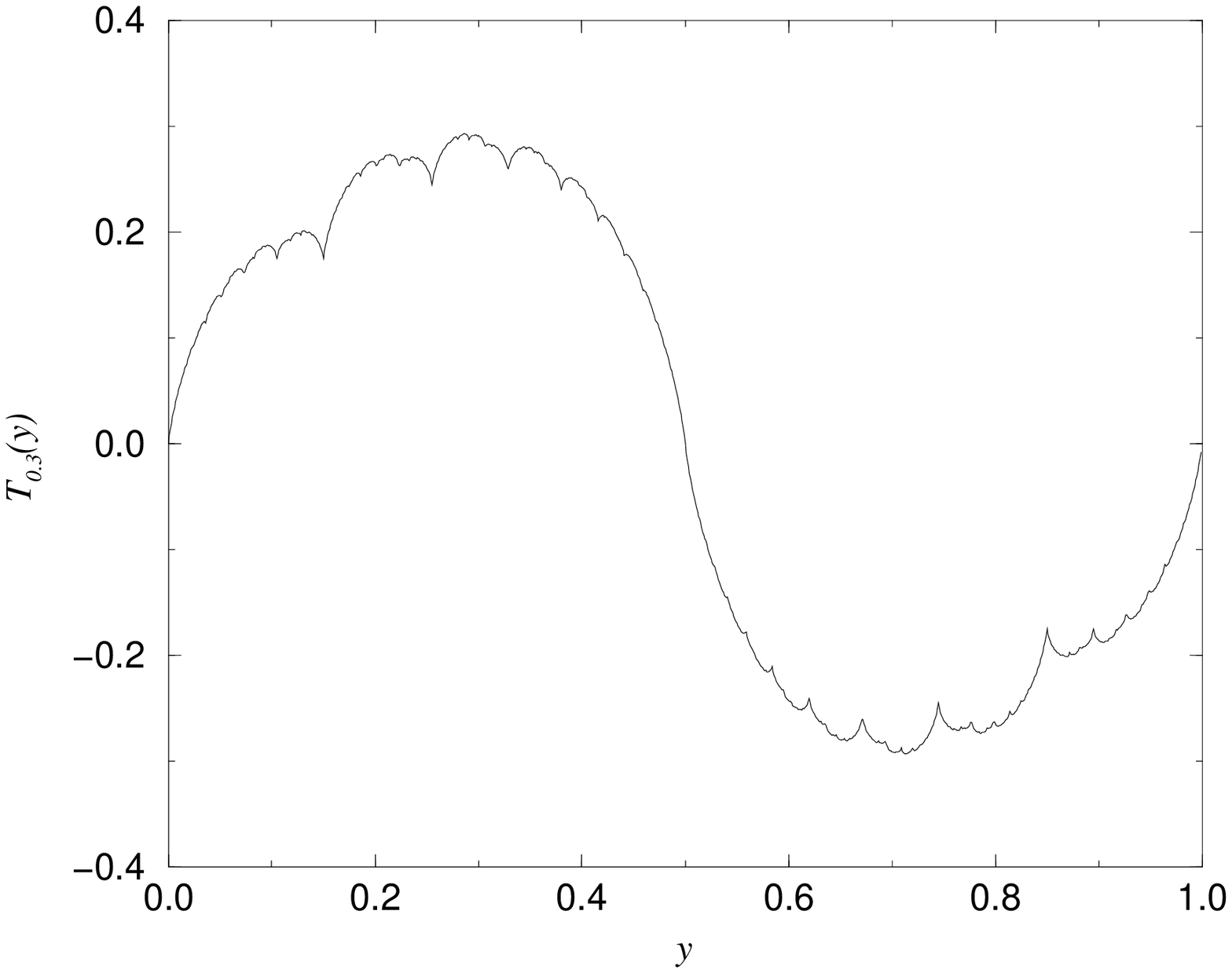,width=8cm}}
\centerline{\psfig{figure=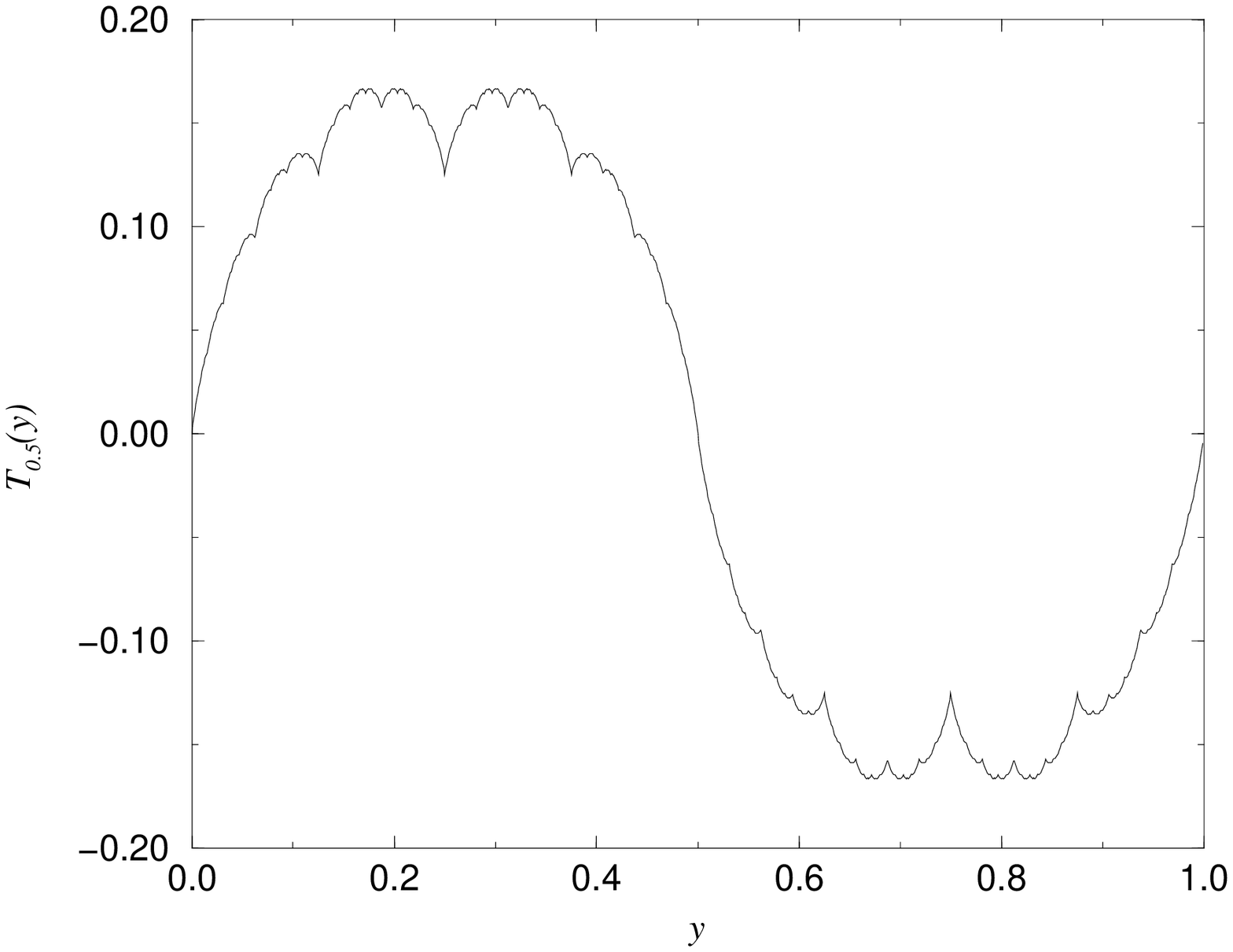,width=8cm}}
\centerline{\psfig{figure=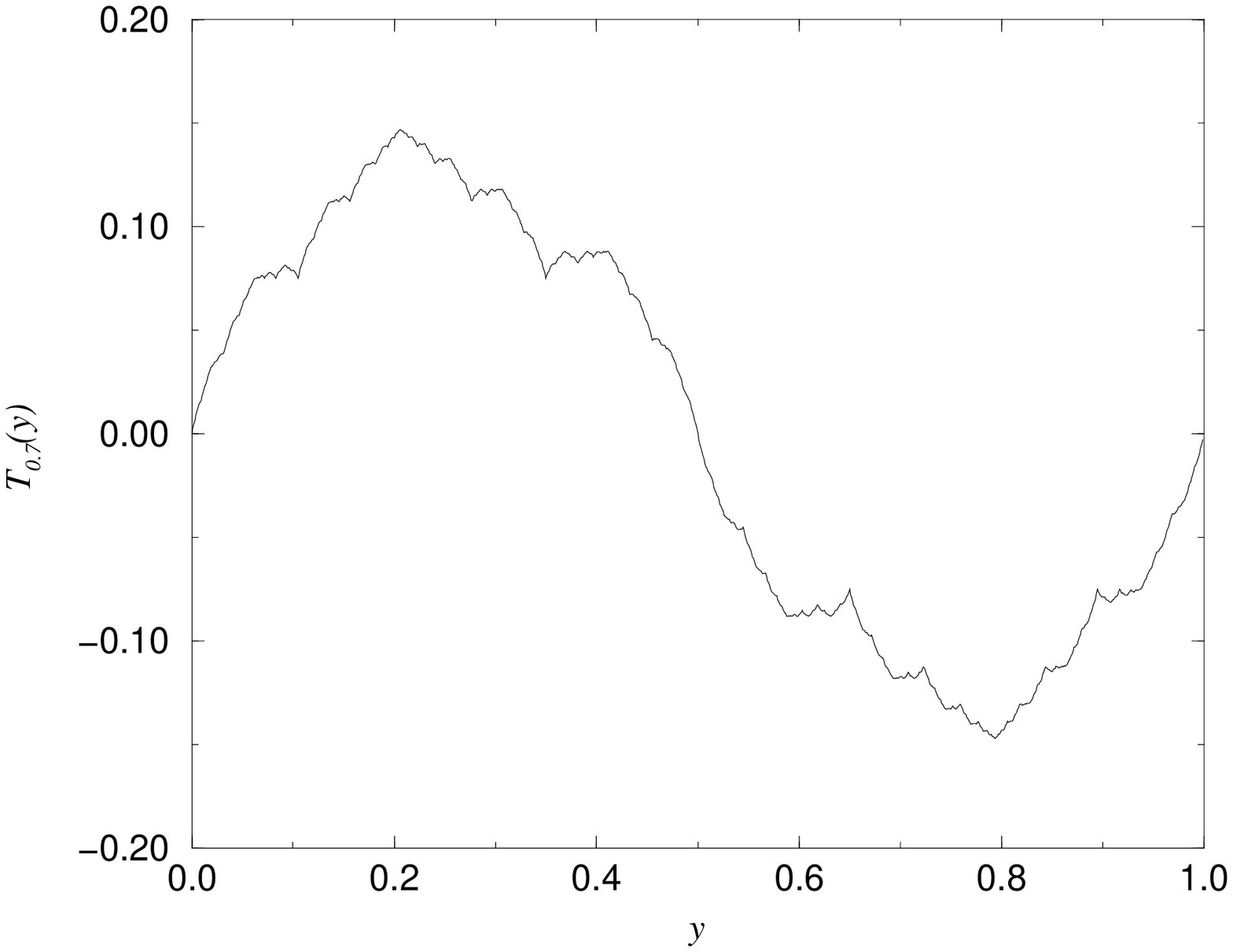,width=8cm}
\psfig{figure=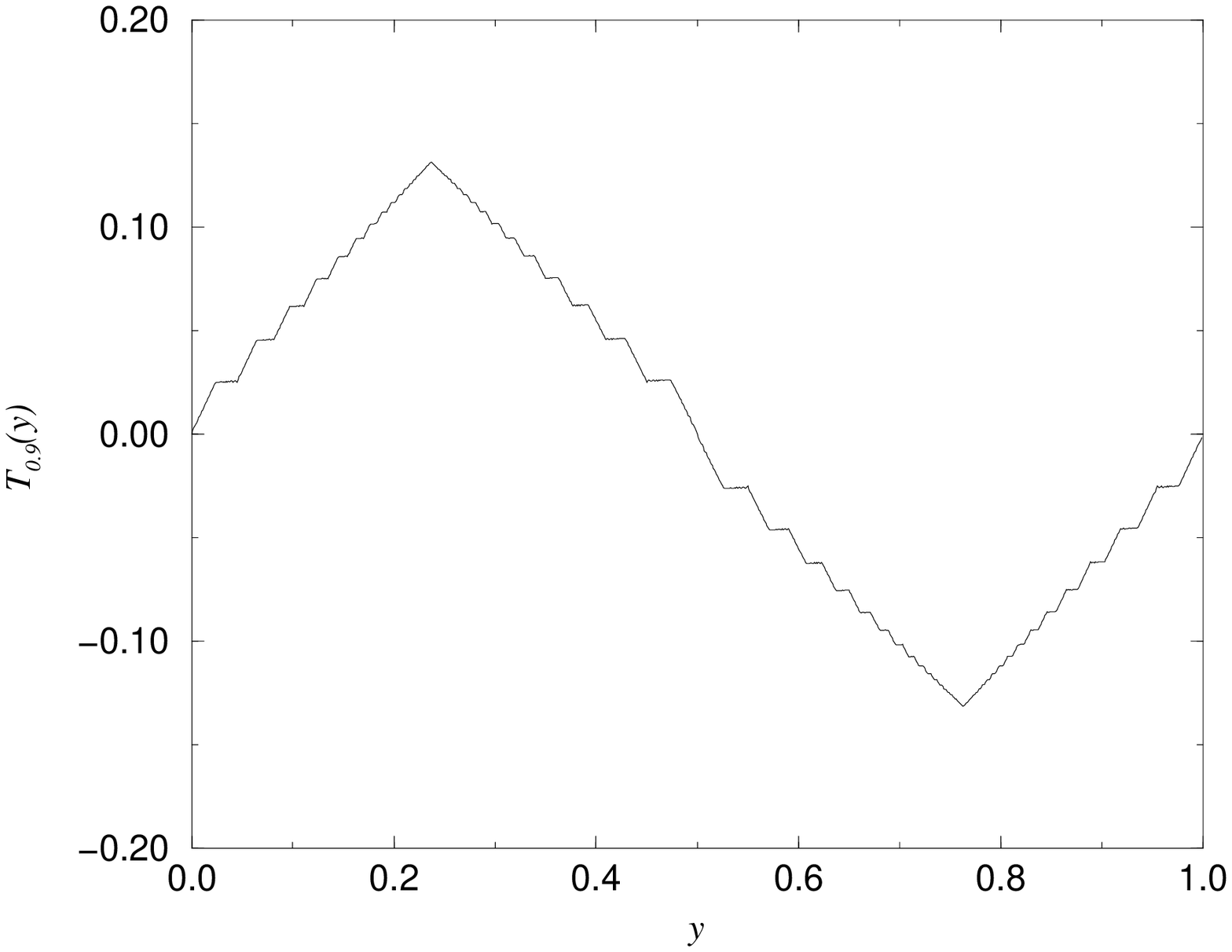,width=8cm}}
\caption{\label{figPRWTqs}{\footnotesize The functions $T_q(y),$ 
Eq. (\ref{PRWtq}), displayed
for five different values of $q = 0.1,\ldots,0.9,$ from top to bottom and 
left to right. The figure in the center,
$q = 0.5,$ corresponds to the symmetric case.}}
\end{figure}

We now want 
to simplify the argument and will make the assumption in the sequel that we
are far enough from the boundaries so that we can ignore the finite size 
effects and replace the incomplete functions $T\p_n$ by their common limit 
value 
\begin{equation}\label{PRWtq}
T_q(y) = 
\left\{
\begin{array}{lr}
\fr{p}{2q} y - q T_q\left(\fr{y}{q} + \fr{1}{2}\right), 
&0 \leq y < \fr{q}{2},\\
\fr{1/2 - y}{2} + p T_q\left(\fr{y - q/2}{p}\right), 
&\fr{q}{2} \leq y < 1 - \fr{q}{2},\\
- \fr{p}{2q} (1 - y) - q T_q\left(\fr{y - 1}{q} + \fr{1}{2}\right), 
&1 - \fr{q}{2} \leq y < 1,
\end{array}
\right.
\end{equation}
%\eqlabel{PRWtq}
where we have made explicit the parametric dependence of the generalized 
Takagi functions $T_q.$ Figure (\ref{figPRWTq}) shows a recursive computation
of $T_q(y),$ for $q = 1/3.$ 
This example is very similar to the Takagi 
function \cite{takagi03}
\begin{equation}
T(y) = 
\left\{
\begin{array}{ll}
y + \fr{1}{2}T(2 y),&0\leq y < \fr{1}{2},\\
1 - y + \fr{1}{2}T(2 y  - 1),&\fr{1}{2}\leq y < 1, 
\end{array}
\right.
\label{takagi}
\end{equation}
%\eqlabel{takagi}
that appears in the example of the open random
walk with flux boundary conditions discussed by Gaspard and co-workers
\cite{tasaki95,gaspard97,gaspard98}.
In particular, for $q = 1/2,$ the case of a symmetric persistent random
walk, up to a factor 2, $T_q$ is identical to $T$
on the first half of the interval and opposite on the second half.
Figure (\ref{figPRWTqs}) shows $T_q(y)$ for different values of $q$ ranging
from 0.1 to 0.9.

\section{\label{IV}Symbolic Dynamics}
%\eqlabel{IV}

In order to compute the entropy production, it is convenient to introduce
a symbolic dynamics. Each half of the unit cell is partitioned into 
sets that correspond to particles that were back- or forward-scattered at
the preceding time step. This defines the 0-partition of a cell,
\begin{equation}\label{PRW0part}
\cA = \{\Gamma^{(-)}_0, \Gamma^{(+)}_0, \Gamma^{(-)}_1, \Gamma^{(+)}_1\},
\end{equation}
%\eqlabel{PRW0part}
where
\begin{equation}\label{PRWGG}
\Gamma\p_i = \Gamma\p \cap \Gamma_i,\: i = 0,1,
\end{equation}
%\eqlabel{PRWGG}
and 
%we further define
\begin{eqnarray}
\label{PRWG-}
\Gamma^{(-)}&=&\left\{(x,y)\::\:0\leq x < \fr{1}{2}\right\},\\
\label{PRWG+}
\Gamma^{(+)}&=&\left\{(x,y)\::\:\fr{1}{2}\leq x < 1\right\},\\
\label{PRWG0}
\Gamma_0&=&\left\{(x,y)\::\:0\leq y < \fr{q}{2}\:{\rm or}\:
1 - \fr{q}{2} \leq y < 1\right\},\\
%\hspace{-6in}{\rm and}&&\nonumber\\
\label{PRWG1}
\Gamma_1 &=& \left\{(x,y)\::\:\fr{q}{2} \leq y < 1 - \fr{q}{2} \right\}.
\end{eqnarray}
%\eqlabel{PRWG-,G+,G0,G1}
An $(l,k)$-partition is the collection of cylinder sets 
\begin{equation}
\Gamma\p_{\omega_{-l},\ldots,\omega_k} = \Gamma\p\cap
\left[\cap_{i=-l}^k \Mprw^i(\Gamma_{\omega_i})\right],
\label{PRWkset}
\end{equation}
%\eqlabel{PRWkset}
where 
$\omega_i \in \{0, 1\},\:i = -l,\ldots, k.$ The measure of 
such a set, $\mu^{(\pm)}_n(\omega_{-l},\ldots,\omega_k),$ can be written,
for $l = 0,$  as
\begin{equation}\label{PRWmeascs}
\Delta g^{(\pm)}_n(\omega_0,\ldots,\omega_k) = 
g^{(\pm)}\Big(n, y(\omega_0,\ldots,\omega_k + 1)\Big) - 
g^{(\pm)}\Big(n, y(\omega_0,\ldots,\omega_k)\Big), 
\end{equation}
%\eqlabel{PRWmeascs}
where $\omega_0,\ldots,\omega_{k - 1}, \omega_k + 1$ is defined by
\begin{equation}
\omega_0, \ldots,\omega_{k - 1}, \omega_k + 1 = 
\left\{
\begin{array}{l@{\quad}l}
\omega_0, \ldots,\omega_{k - 1}, 1,&\omega_k = 0,\\
\omega_0, \ldots,\omega_{k - 1} + 1,0&\omega_k = 1,
\end{array}
\right.
\label{omegap1}
\end{equation}
%\eqlabel{omegap1}
with the further convention that $y(1,\ldots,1, 1 + 1) = 1.$

By looking at the $y$-components of $\Mprw$ in Eq. (\ref{PRWmap}), we find
that $y(\omega_0,\ldots, \omega_k)$ can be computed through the recursion
relation
\begin{equation}\label{PRWyk}
y(\omega_0,\ldots, \omega_k) = \left(\nu(\omega_0)y(\omega_1,\ldots,
\omega_k) - (-)^{\omega_0}\fr{q}{2}\right) \mod 1,
\end{equation}
%\eqlabel{PRWyk}
where we set
\begin{equation}\label{PRWnu}
\nu(\omega_0) = 
\left\{
\begin{array}{lr}
q,\quad&\omega_0 = 0,\\
p,\quad&\omega_0 = 1.
\end{array}
\right.
\end{equation}
%\eqlabel{PRWnu}

Therefore the functions $T_q$ in Eq. (\ref{PRWtq}) can be defined directly 
in terms of the symbolic sequences,
\begin{equation}\label{PRWTq}
T_q(\omega_0,\ldots,\omega_k) = 
\left\{
\begin{array}{lr}
\fr{p}{2}\left(y(\omega_1,\ldots,\omega_k) - \fr{1}{2}\right)
- qT_q(\omega_1,\ldots,\omega_k),\quad&\omega_0 = 0,\\
-\fr{p}{2}\left(y(\omega_1,\ldots,\omega_k) - \fr{1}{2}\right)
+ pT_q(\omega_1,\ldots,\omega_k),\quad&\omega_0 = 1.
\end{array}
\right.
\end{equation}
%\eqlabel{PRWTq}

With the help of Eq. (\ref{PRWsolcf}), we can then rewrite Eq. 
(\ref{PRWmeascs})
as
\begin{equation}\label{PRWmeascs2}
\Delta g^{(\pm)}_n(\omega_0,\ldots,\omega_k) = 
\left(\fr{n + 1}{2} \mp \fr{1}{4q}\right)\Delta y(\omega_0,\ldots,\omega_k)
\pm \Delta T_q(\omega_0,\ldots,\omega_k),
\end{equation}
%\eqlabel{PRWmeascs2}
with obvious notations for $\Delta y$ and $\Delta T_q.$ We note here that
our assumption that the cell $n$ is sufficiently far away from the boundaries 
means in terms of the size of the cylinder sets that $k$ must be
strictly less than the distance to the closest boundary. In that case, we
can substitute without loss of generality $T\p_n$ by $T_q$ in 
Eq. (\ref{PRWsolcf}).

For later purposes, we note that, from Eqs. (\ref{PRWyk}, \ref{PRWTq}), we 
have
\begin{equation}\label{PRWDy}
\Delta y(\omega_0,\ldots,\omega_k) = 
\nu(\omega_0)\Delta y(\omega_1,\ldots,\omega_k) 
= \prod_{i = 0}^k \nu(\omega_i),
\end{equation}
%\eqlabel{PRWDy}
and
\begin{equation}\label{PRWDTq}
\Delta T_q(\omega_0,\ldots,\omega_k) = 
\left\{
\begin{array}{lr}
\fr{p}{2}\Delta y(\omega_1,\ldots,\omega_k)
- q\Delta T_q(\omega_1,\ldots,\omega_k),\quad&\omega_0 = 0,\\
-\fr{p}{2}\Delta y(\omega_1,\ldots,\omega_k)
+ p\Delta T_q(\omega_1,\ldots,\omega_k),\quad&\omega_0 = 1.
\end{array}
\right.
\end{equation}
%\eqlabel{PRWDTq}

\section{{\it \lowercase{k}}-entropy and 
entropy production rate\label{V}}
%\eqlabel{V}

Following \cite{gilbert99b}, we write the entropy of a $(l, k)$-partition as 
\begin{equation}\label{PRWSk}
S_{l,k}(n) \equiv S^{(-)}_{l,k}(n) + S^{(+)}_{l,k}(n),
\end{equation}
%\eqlabel{PRWSk}
and
\begin{equation}\label{PRWSlkpm}
S_{l,k}^{(\pm)}(n) = - \sum_{\omega_{-l},\ldots,\omega_{k - 1}}
\mu\p_n(\omega_{-l},\ldots,\omega_{k - 1})\left[
\log\fr{\mu\p_n(\omega_{-l},\ldots,\omega_{k - 1})}
{\nu\p(\omega_{-l},\ldots,\omega_{k - 1})} - 1\right],
\end{equation}
%\eqlabel{PRWSlkpm}

Because
the invariant measure is uniform along the expanding direction, 
the ratio of $\mu\p_n$ and $\nu\p$ obeys the identity
\begin{equation}
\fr{\mu\p_n(\omega_{-l},\ldots,\omega_{k - 1})}
{\nu\p(\omega_{-l},\ldots,\omega_{k - 1})}
= \fr{\mu\p_n(\omega_{0},\ldots,\omega_{k - 1})}
{\nu\p(\omega_{0},\ldots,\omega_{k - 1})}.
\label{ratiomunu}
\end{equation}
%\eqlabel{ratiomunu}
Therefore the entropy is extensive with respect
to the $x$-direction and we have
\begin{equation}
S\p_{l, k}(n) = S\p_{0, k}(n).
\label{PRWlkentext}
\end{equation}
%\eqlabel{PRWlkentext}
Henceforth, we will drop the $l$ dependence and will simply consider the 
$k$-entropies, which we write
\begin{equation}\label{PRWSkpm}
S_{k}^{(\pm)}(n) = - \sum_{\uo{k}}\Delta g\p_n(\uo{k})
\left[\log\fr{\Delta g\p_n(\uo{k})}{\nu\p(\uo{k})} - 1\right],
\end{equation}
%\eqlabel{PRWSkpm}
where we used the compact notation $\uo{k} \equiv 
\omega_0,\ldots,\omega_{k - 1}$. Here $\nu\p(\uo{k})$ is the
volume of the corresponding cylinder set,
\begin{equation}\label{PRWnuk}
\nu\p(\uo{k}) = \fr{1}{2} \Delta y(\uo{k})
= \fr{1}{2}\prod_{i = o}^{k - 1}\nu(\omega_i).
\end{equation}
%\eqlabel{PRWnuk}

The entropy production rate follows by a straightforward generalization
of the formalism detailed in \cite{gilbert99b} to Eq. (\ref{PRWSkpm})~:
\begin{equation}\label{PRWDiSk}
\Delta_i S_k(n) \equiv \Delta_i S^{(-)}_k(n) + \Delta_i S^{(+)}_k(n),
\end{equation}
%\eqlabel{PRWDiSk}
and
\begin{equation}\label{PRWDiSkpm}
\Delta_i S_k^{(\pm)}(n) = \sum_{\uo{k + 1}}\Delta g^{(\pm)}_n(\uo{k + 1})
\log\fr{\Delta g^{(\pm)}_n(\uo{k + 1})}{\nu(\omega_k)
\Delta g^{(\pm)}_n(\uo{k})}.
\end{equation}
%\eqlabel{PRWDiSkpm}

Assuming that the stationary state is dominated by the linear part,
we can compute the $k$-entropy production rate by expanding the expressions
for $\Delta_i S_k^{(\pm)}(n)$ in Eq.~(\ref{PRWDiSkpm}) in powers of 
\begin{equation}\label{PRWratio}
\fr{\Delta T_q}{(\fr{n + 1}{2} \pm \fr{1}{4q})\Delta y},
\end{equation}
%\eqlabel{PRWratio}
which we further expand in powers of $\fr{1}{n + 1}.$ To first order, 
Eq.~(\ref{PRWDiSk}) becomes
\begin{equation}\label{PRWexpDiSk}
\Delta_i S_k(n) = \fr{2}{n + 1}\sum_{\uo{k + 1}}
\fr{[\Delta T_q(\uo{k + 1}) - \nu(\omega_k)\Delta T_q(\uo{k})]^2}
{\Delta y(\uo{k + 1})}.
\end{equation}
%\eqlabel{PRWexpDiSk}
Making use of Eqs.~(\ref{PRWDy}-\ref{PRWDTq}), we readily see that this 
expression is independent of $k,$ i.e.
\begin{eqnarray}
\lefteqn{\sum_{\uo{k + 1}}\fr{[\Delta T_q(\uo{k + 1}) - 
\nu(\omega_k)\Delta T_q(\uo{k})]^2}
{\Delta y(\uo{k + 1})}}\nonumber\\ 
&=&\sum_{\uo{k}}\fr{[\Delta T_q(\uo{k}) - \nu(\omega_{k - 1})
\Delta T_q(\uo{k - 1})]^2}{\Delta y(\uo{k})}\nonumber\\
&=& \sum_{\omega = 0,1} \fr{[\Delta T_q(\omega)]^2}{\Delta y(\omega)}
\nonumber\\
&=& \fr{p}{4q}.\label{PRWsumk}
\end{eqnarray}
%\eqlabel{PRWsumk}
Therefore,
\begin{equation}\label{PRWDiSk1}
\Delta_iS_k(n) = \fr{p}{2q} \fr{1}{n + 1} + O\left(\fr{1}{(n + 1)^3}\right).
\end{equation}
%\eqlabel{PRWDiSk1}

The derivation of the next order term proceeds as follows.
For the sake of simplifying the notations, we will write 
Eq.~(\ref{PRWy1cf}) as
\begin{equation}\label{PRWy1cf2}
\gpm \equiv g^{(\pm)}(n, 1)
\end{equation}
%\eqlabel{PRWy1cf2}

We start from Eq.~(\ref{PRWDiSkpm}) and substitute Eq.~(\ref{PRWmeascs2}) for
$\Delta g^{(\pm)}_n(\uo{k + 1})$ :
\begin{eqnarray}
\Delta_i S_k^{(\pm)}(n) &=& \sum_{\uo{k + 1}}\Delta g^{(\pm)}_n(\uo{k} + 1)
\log\fr{\Delta g^{(\pm)}_n(\uo{k + 1})}{\nu(\omega_k)
\Delta g^{(\pm)}_n(\uo{k})},
\nonumber\\
&=& \sum_{\uo{k + 1}}\left[\gpm
\Delta y(\uo{k + 1}) \pm \Delta T_q(\uo{k + 1})\right]
\nonumber\\
&&\phantom{\sum}\log \fr{\left[\gpm
\Delta y(\uo{k + 1}) \pm \Delta T_q(\uo{k + 1})\right]}
{\left[\gpm
\Delta y(\uo{k + 1}) \pm \nu(\omega_k)\Delta T_q(\uo{k})\right]},
\label{PRWDiSkpm2}
\end{eqnarray}
%\eqlabel{PRWDiSkpm2}
where, in the last line, we used
\begin{equation}\label{PRWprop1}
\nu(\omega_k)\Delta y(\uo{k}) = \Delta y(\uo{k + 1}).
\end{equation}
%\eqlabel{PRWprop1}

Factoring $\gpm\Delta y(\uo{k + 1})$ in the logarithms and expanding up to
fourth order, Eq.(\ref{PRWDiSkpm2}) becomes
\begin{eqnarray}
\lefteqn{
\sum_{\uo{k + 1}}\left[\gpm
\Delta y(\uo{k + 1}) \pm \Delta T_q(\uo{k + 1})\right]}\nonumber\\
&&\phantom{\sum}\left\{
\log \left[1 \pm \fr{\Delta T_q(\uo{k + 1})}{\gpm\Delta y(\uo{k + 1})}\right]
- \log \left[1 \pm \fr{\nu(\omega_k)\Delta T_q(\uo{k})}
{\gpm\Delta y(\uo{k + 1})}\right]
\right\} =\nonumber\\
&&\sum_{\uo{k + 1}}\left[\gpm
\Delta y(\uo{k + 1}) \pm \Delta T_q(\uo{k + 1})\right]
\Bigg\{
\pm \fr{\Delta T_q(\uo{k + 1})}{\gpm\Delta y(\uo{k + 1})} 
\nonumber\\
&&\mp \fr{\nu(\omega_k)\Delta T_q(\uo{k})}
{\gpm\Delta y(\uo{k + 1})}
- \fr{\Delta T_q(\uo{k + 1})^2}{2\left[\gpm\Delta y(\uo{k + 1})\right]^2}
\nonumber\\
&&
+ \fr{\nu(\omega_k)^2\Delta T_q(\uo{k})^2}
{2\left[\gpm\Delta y(\uo{k + 1})\right]^2}
\pm \fr{\Delta T_q(\uo{k + 1})^3}{3\left[\gpm\Delta y(\uo{k + 1})\right]^3}
\mp \fr{\nu(\omega_k)^3\Delta T_q(\uo{k})^3}
{3\left[\gpm\Delta y(\uo{k + 1})\right]^3}
\nonumber\\
&&
- \fr{\Delta T_q(\uo{k + 1})^4}{4\left[\gpm\Delta y(\uo{k + 1})\right]^4}
+ \fr{\nu(\omega_k)^4\Delta T_q(\uo{k})^4}
{4\left[\gpm\Delta y(\uo{k + 1})\right]^4}
\Bigg\},\label{PRWDiSkpm3}
\end{eqnarray}
%\eqlabel{PRWDiSkpm3}
which, keeping terms up to $O(1/(n + 1)^3),$ takes the form
\begin{eqnarray}
\lefteqn{\sum_{\uo{k + 1}}\Bigg\{\pm \Delta T_q(\uo{k + 1}) \mp \nu(\omega_k)
\Delta T_q(\uo{k}) + \fr{\Delta T_q(\uo{k + 1})^2}{2\gpm\Delta y(\uo{k + 1})}}
\nonumber\\
&& -\fr{\nu(\omega_k)\Delta T_q(\uo{k + 1})\Delta T_q(\uo{k})}
{\gpm\Delta y(\uo{k + 1})}
+\fr{\nu(\omega_k)^2\Delta T_q(\uo{k})^2}{2\gpm\Delta y(\uo{k + 1})}
\nonumber\\
&& \mp \fr{\Delta T_q(\uo{k + 1})^3}
{6\left[\gpm\Delta y(\uo{k + 1})\right]^2}
\pm \fr{\nu(\omega_k)^2\Delta T_q(\uo{k + 1})\Delta T_q(\uo{k})^2}
{2\left[\gpm\Delta y(\uo{k + 1})\right]^2}
\nonumber\\
&&
\mp  \fr{\nu(\omega_k)^3\Delta T_q(\uo{k})^3}
{3\left[\gpm\Delta y(\uo{k + 1})\right]^2}
+ \fr{\Delta T_q(\uo{k + 1})^4}{12\left[\gpm\Delta y(\uo{k + 1})\right]^3}
\nonumber\\
&&
- \fr{\nu(\omega_k)^3\Delta T_q(\uo{k + 1})\Delta T_q(\uo{k})^3}
{3\left[\gpm\Delta y(\uo{k + 1})\right]^3}
+ \fr{\nu(\omega_k)^4\Delta T_q(\uo{k})^4}
{4\left[\gpm\Delta y(\uo{k + 1})\right]^3}
\Bigg\}.\label{PRWDiSkpm4}
\end{eqnarray}
%\eqlabel{PRWDiSkpm4}

Therefore, the $k$-entropy production rate, Eq.(\ref{PRWDiSk}), reads
\begin{eqnarray}
\lefteqn{\Delta_i S_k(n) =\left(\fr{1}{2\gp} + \fr{1}{2\gm}\right)
\sum_{\uo{k + 1}}
\fr{\left[\Delta T_q(\uo{k + 1}) - \nu(\omega_k)\Delta T_q(\uo{k})\right]^2}
{\Delta y(\uo{k + 1})}}
\nonumber\\
&& + \left(\fr{1}{6\gp^2} - \fr{1}{6\gm^2}\right)\nonumber\\
&&\sum_{\uo{k + 1}} \fr{- \Delta T_q(\uo{k + 1})^3 + 3\nu(\omega_k)^2
\Delta T_q(\uo{k + 1}) \Delta T_q(\uo{k})^2 - 2\nu(\omega_k)^3
\Delta T_q(\uo{k})^3}
{\Delta y(\uo{k + 1})^2}
\nonumber\\
&& + \left(\fr{1}{12\gp^3} + \fr{1}{12\gm^3}\right)\nonumber\\
&&\sum_{\uo{k + 1}} \fr{\Delta T_q(\uo{k + 1})^4 - 4\nu(\omega_k)^3
\Delta T_q(\uo{k + 1}) \Delta T_q(\uo{k})^3 + 3\nu(\omega_k)^4
\Delta T_q(\uo{k})^4}
{\Delta y(\uo{k + 1})^3}.\nonumber\\
{ }\label{PRWDiSk2}
\end{eqnarray}
%\eqlabel{PRWDiSk2}
Assuming that $n$ is large, we can expand the ratios involving $\gpm$
in powers of $1/(n + 1).$ To third order, we get
\begin{eqnarray}
&&\fr{1}{2\gp} + \fr{1}{2\gm} = \fr{2}{n + 1} + \fr{1}{2q^2(n + 1)^3},
\label{PRWratio1}\\
&&\fr{1}{6\gp^2} - \fr{1}{6\gm^2} = \fr{4}{3q(n + 1)^3},\label{PRWratio2}\\
&&\fr{1}{12\gp^3} + \fr{1}{12\gm^3} = \fr{4}{3(n + 1)^3}.\label{PRWratio3}
\end{eqnarray}
%\eqlabel{PRWratio1-3}

Therefore, to leading order in $1/(n + 1),$ we retrieve 
Eq.~(\ref{PRWexpDiSk}), which, along with Eq.~(\ref{PRWsumk}), gives 
Eq.(\ref{PRWDiSk1}).

To work out the next order terms, we first note, with the help of 
Eq.(\ref{PRWprop1}), that expressions involving both $\Delta T_q (\uo{k + 1})$ 
and $\Delta T_q (\uo{k})$ can be transformed into expressions involving only 
$\Delta T_q (\uo{k})$ by summing over $\omega_k.$ For instance,
\begin{equation}\label{PRWprop2}
\fr{\nu(\omega_k)^2\Delta T_q(\uo{k + 1})\Delta T_q(\uo{k})^2}
{\Delta y(\uo{k + 1})^2} = \fr{\Delta T_q(\uo{k + 1})\Delta T_q(\uo{k})^2}
{\Delta y(\uo{k})^2}.
\end{equation}
%\eqlabel{PRWprop2}
Now, by definition,
\begin{eqnarray}
\sum_{\omega_k} \Delta T_q(\uo{k + 1}) &=& \sum_{\omega_k} \left[T_q(\uo{k}, 
\omega_k + 1) - T_q(\uo{k}, \omega_k)\right],\nonumber\\
&=& T_q(\uo{k - 1}, \omega_{k - 1} + 1) - T_q(\uo{k}),\nonumber\\
&=& \Delta T_q(\uo{k}).
\label{PRWprop3}
\end{eqnarray}
%\eqlabel{PRWprop3}
Therefore,
\begin{equation}\label{PRWprop4}
\sum_{\omega_k}\fr{\nu(\omega_k)^2\Delta T_q(\uo{k + 1})\Delta T_q(\uo{k})^2}
{\Delta y(\uo{k + 1})^2} = \fr{\Delta T_q(\uo{k})^3}{\Delta y(\uo{k})^2},
\end{equation}
%\eqlabel{PRWprop4}
and, similarly,
\begin{equation}\label{PRWprop5}
\sum_{\omega_k}\fr{\nu(\omega_k)^3\Delta T_q(\uo{k + 1})\Delta T_q(\uo{k})^3}
{\Delta y(\uo{k + 1})^3} = \fr{\Delta T_q(\uo{k})^4}{\Delta y(\uo{k})^3}.
\end{equation}
%\eqlabel{PRWprop5}
 
Combining Eqs.~(\ref{PRWratio1}-\ref{PRWratio2}, 
\ref{PRWprop2}-\ref{PRWprop5}),
we can rewrite the last two lines of Eq.~(\ref{PRWDiSk2}) as
\begin{eqnarray}
&&-\fr{4}{3q(n + 1)^3}\left[\sum_{\uo{k + 1}}\fr{\Delta T_q(\uo{k + 1})^3}
{\Delta y(\uo{k + 1})^2} - \sum_{\uo{k}}\fr{\Delta T_q(\uo{k})^3}
{\Delta y(\uo{k})^2}\right] \nonumber\\
&&+ \fr{4}{3(n + 1)^3}
\left[\sum_{\uo{k + 1}}\fr{\Delta T_q(\uo{k + 1})^4}
{\Delta y(\uo{k + 1})^3} - \sum_{\uo{k}}\fr{\Delta T_q(\uo{k})^4}
{\Delta y(\uo{k})^3}\right].
\label{PRWDiSk3}
\end{eqnarray}
%\eqlabel{PRWDiSk3}

We can work out these expressions by substituting Eq.(\ref{PRWDTq}) for
$\Delta T_q(\uo{k + 1})$ and summing over $\omega_0.$ Before doing so, we will
need the following
\begin{eqnarray}
\sum_{\uo{k + 1}} \Delta y(\uo{k + 1}) &=& 1,\label{PRWprop6}\\
\sum_{\uo{k + 1}} \Delta T_q (\uo{k + 1}) &=& \sum_{\omega} 
\Delta T_q(\omega) = 0,\label{PRWprop7}\\
\sum_{\uo{k + 1}} \fr{\Delta T_q (\uo{k + 1})^2}{\Delta y(\uo{k + 1})}&=&
\fr{\left[\fr{p}{2}\Delta y(\uo{k}) - q \Delta T_q(\uo{k})\right]^2}
{q\Delta y(\uo{k})}\nonumber\\
&&+ \fr{\left[\fr{p}{2}\Delta y(\uo{k}) - 
p \Delta T_q(\uo{k})\right]^2}
{p\Delta y(\uo{k})},\nonumber\\
&=& \fr{p^2}{4}\left(\fr{1}{q} + \fr{1}{p}\right) + \sum_{\uo{k}}
\fr{\Delta T_q(\uo{k})^2}{\Delta y(\uo{k})},\nonumber\\
&=& (k + 1)\fr{p}{4q}.\label{PRWprop8}
\end{eqnarray}
%\eqlabel{PRWprop6-8}
For the cubic term, we
have
\begin{eqnarray}
\sum_{\uo{k + 1}}\fr{\Delta T_q(\uo{k + 1})^3}{\Delta y(\uo{k + 1})^2}
&=& \sum_{\uo{k}}
\fr{\left[\fr{p}{2}\Delta y(\uo{k}) - q \Delta T_q(\uo{k})\right]^3}
{q^2\Delta y(\uo{k})^2} \nonumber\\
&&- \fr{\left[\fr{p}{2}\Delta y(\uo{k}) 
- p \Delta T_q(\uo{k})\right]^3}
{p^2\Delta y(\uo{k})^2},\nonumber\\
&=& \fr{p^3}{8}\left(\fr{1}{q^2} - \fr{1}{p^2}\right) 
+ (p - q)\sum_{\uo{k}}
\fr{\Delta T_q(\uo{k})^3}{\Delta y(\uo{k})^2},\nonumber\\
&=& \fr{p}{8 q^2}\sum_{i = 1}^{k} (p - q)^i
= \fr{p}{16 q^3}(p - q)\left[1 - (p - q)^k\right],\nonumber\\
{}\label{PRWprop9}
\end{eqnarray}
%\eqlabel{PRWprop9}
and, for the quartic term,
\begin{eqnarray}
\sum_{\uo{k + 1}}\fr{\Delta T_q(\uo{k + 1})^4}{\Delta y(\uo{k + 1})^3}
&=& \sum_{\uo{k}}
\fr{\left[\fr{p}{2}\Delta y(\uo{k}) - q \Delta T_q(\uo{k})\right]^4}
{q^3\Delta y(\uo{k})^3} \nonumber\\
&&+ 
\fr{\left[\fr{p}{2}\Delta y(\uo{k}) - p \Delta T_q(\uo{k})\right]^4}
{p^3\Delta y(\uo{k})^3},\nonumber\\
&=& \fr{p^4}{16}\left(\fr{1}{q^3} + \fr{1}{p^3}\right)
+ k \fr{3p^2}{8q^2}
- \fr{p^2}{4 q^3}(p - q)\left[1 - (p - q)^{k - 1}\right]
\nonumber\\
&&+ \sum_{\uo{k}}\fr{\Delta T_q (\uo{k})^4}{\Delta y(\uo{k})^3}.
\label{PRWprop10}
\end{eqnarray}
%\eqlabel{PRWprop10}

Equation (\ref{PRWDiSk2}) combined with 
Eqs.~(\ref{PRWDiSk2}, \ref{PRWratio1}-\ref{PRWratio3}, 
\ref{PRWDiSk3}, \ref{PRWprop9}-\ref{PRWprop10}) yield the third order 
correction to the entropy production rate~:
\begin{eqnarray}
\Delta_i S_k(n) &=& \fr{p}{2q}\fr{1}{n + 1} + 
\left[\fr{p}{q^3}\left(\fr{1}{8} - \fr{p^2}{3} + \fr{p^3}{12}\right)\right.
\nonumber\\
&&\left.
+ \fr{p}{12} + \fr{p^2}{3q^2} + \fr{p^2}{2q^2}k + (p - q)^{k + 1}\fr{p}{6q^3}
\right]\fr{1}{(n + 1)^3}\nonumber\\
&&+ O\left(\fr{1}{(n + 1)^5}\right).\label{PRWkep}
\end{eqnarray} 
%\eqlabel{PRWkep}

We note that, in the symmetric case, $p = q = 1/2,$ the coefficient in front
of the third order term is $7/12 + k/2.$ The term linear in $k$ is identical
to the case of the random walk \cite{gaspard97,gaspard98}, but the first 
term is different. For 
$q \neq p,$ we find a new term, proportional to $(p - q)^k,$ which decays
exponentially. Thus, for large $k,$ Eq.~(\ref{PRWkep}) is at least 
qualitatively similar to the case of the multi-baker map with a linear 
divergence in $k$ that is third order in the small parameter $1/(n + 1).$

\section{\label{VI}Discussion}

We have shown that the entropy production formalism of Gaspard, Gilbert
and Dorfman applies successfully to a persistent random walk driven 
away from equilibrium by a density current. The leading order term in 
the expansion in inverse powers of the spatial coordinate is in 
exact agreement with the phenomenological entropy production, 
Eq. (\ref{OVssentprod}). It is particularly important to note that this term is independent of the
coarse-graining parameter, $k$. Our result is therefore similar to Gaspard's
for the simpler case of a random walk \cite{gaspard97,gaspard98}, and
we also find
\begin{equation}
\lim_{k\rightarrow\infty}\lim_{|\vec{\nabla} \mu_n|/\mu_n\rightarrow 0}
\lim_{L\rightarrow\infty} \fr{\mu_n}{(\vec{\nabla} \mu_n)^2}\Delta_iS_k = D,
\label{PRWlimep}
\end{equation}
%\eqlabel{PRWlimep}
where we wrote $\mu_n = g(n, 1)$ and $\vec{\nabla}$ denotes the density 
gradient with respect to the lattice coordinate.
The limit on the resolution parameter is taken last, which is to say that
the resolution dependent terms are accounted for by finite size effects
and have thus no counterpart in thermodynamics, where one assumes that
systems have infinitely many degrees of freedom.

It is interesting to note that the form of the first order contribution
to the $k$-entropy production rate, Eq. (\ref{PRWexpDiSk}), is rather 
universal. Indeed similar expressions arise in other baker map models
\cite{gaspard97,gilbert99b,tasaki99}.
It would therefore be interesting to understand  in a more general setting 
the connection between the form of the generalized Takagi functions and the 
expression of the entropy production involving the diffusion coefficient. 
This question seems to be
at the heart of the agreement between the dynamical and phenomenological
approaches to entropy production and still needs further explanation.

\begin{acknowledgements}
The authors wish to thank P. Gaspard and S. Tasaki as well as M. H. Ernst, 
E. G. D. Cohen, L. Rondoni, J. Vollmer, and R. Klages for interesting and 
fruitful discussions. J. R. D. wishes to acknowledge
support from the National Science Foundation under grant PHY 96-00428.
\end{acknowledgements}

%\eqlabel{figPRWvmap}

\newpage
%\eqlabel{figPRWmap}

\newpage
%\eqlabel{figPRWTMTM}

\newpage
%\eqlabel{figPRWSMSM}

\newpage
%\eqlabel{figPRWTS}

\newpage
%\eqlabel{figPRWincT}

\newpage
%\eqlabel{figPRWTq}

\newpage
%\eqlabel{figPRWTqs}

\end{document}